\documentclass[sigconf, nonacm]{acmart}

\copyrightyear{2026}
\acmYear{2026}
\setcopyright{none}
\setcctype{by}
\acmConference[CHI '26]{Proceedings of the 2026 CHI Conference on Human Factors in Computing Systems}{April 13--17, 2026}{Barcelona, Spain}
\acmBooktitle{Proceedings of the 2026 CHI Conference on Human Factors in Computing Systems (CHI '26), April 13--17, 2026, Barcelona, Spain}
\acmPrice{}
\acmDOI{10.1145/3772318.3790439}
\acmISBN{979-8-4007-2278-3/2026/04}

\begin{document}

\title{Reporting and Reviewing LLM-Integrated Systems in HCI: Challenges and Considerations} %

\author{Karla Felix Navarro}
\affiliation{%
  \institution{Department of Computer Science and Operations Research (DIRO)}
  \institution{Université de Montréal}
  \city{Montreal}
  \state{Quebec}
  \country{Canada}
}
\email{karla.felix.navarro@umontreal.ca}

\author{Eugene Syriani}
\affiliation{%
  \institution{Department of Computer Science and Operations Research (DIRO)}
  \institution{Université de Montréal}
  \city{Montreal}
  \state{Quebec}
  \country{Canada}
}
\email{syriani@iro.umontreal.ca}

\author{Ian Arawjo}
\affiliation{%
  \institution{Department of Computer Science and Operations Research (DIRO)}
  \institution{Université de Montréal}
  \city{Montreal}
  \state{Quebec}
  \country{Canada}
}
\email{ian.arawjo@umontreal.ca}

\renewcommand{\shortauthors}{Navarro, Syriani, and Arawjo}

\begin{abstract}

What should HCI scholars consider when reporting and reviewing papers that involve LLM-integrated systems? We interview 18 authors of LLM-integrated system papers on their authoring and reviewing experiences. We find that norms of trust-building between authors and reviewers appear to be eroded by the uncertainty of LLM behavior and hyperbolic rhetoric surrounding AI. %
Authors perceive that reviewers apply uniquely skeptical and inconsistent standards towards papers that report LLM-integrated systems, and mitigate mistrust by adding technical evaluations, justifying usage, and de-emphasizing LLM presence. Authors' views  challenge blanket directives to report all prompts and use open models, arguing that prompt reporting is context-dependent and justifying proprietary model usage despite ethical concerns.
Finally, some tensions in peer review appear to stem from clashes between the norms and values of HCI and ML/NLP communities, particularly around what constitutes a contribution and an appropriate level of technical rigor. 
Based on our findings and additional feedback from six expert HCI researchers, we present a set of considerations for authors, reviewers, and HCI communities around reporting and reviewing papers that involve LLM-integrated systems. Proposed reporting guidelines are also presented on our \href{https://ianarawjo.github.io/Guidelines-for-Reporting-LLM-Integrated-Systems-in-HCI/}{\textcolor{blue}{companion website}}.

\end{abstract}




\maketitle

\section{Introduction}

\begin{quote}
    ``We will add all prompts to an Appendix.'' \phantom{~} \hfill---an HCI system paper author, arguing in a rebuttal for their paper's acceptance
\end{quote}

Across HCI and computing conferences at large, we are seeing an influx of papers that involve large language models (LLMs) \cite{movva2024topics, pangLLMificationCHI2025}. This trend has taken on a reflexive character---papers, such as this one, typically begin with some variant on ``with the rise of LLMs,'' and include modifiers like ``LLM-powered'' in their titles and abstracts. In HCI research, software systems submitted to conferences increasingly integrate LLMs, whether as one-off support for parts of the interface, or as central features. %
We call these ``LLM-integrated systems.''  %

Amid this rise, we, as a community, have not had time to introspect and %
clarify our standards and methodological commitments when reporting and reviewing LLM-integrated systems. %
On the one hand, authors can be uncertain how best to report these systems, and may either underfit their reporting to fit within word limits (omit critical details like prompts, models, architecture, or engineering methodologies), or overfit, burying contributions in unnecessary complexity (include dozens of prompts in expansive appendices). On the other hand, reviewers of HCI papers, in the absence of guidelines around reporting, might feel emboldened to make unsubstantiated claims around LLMs or hold authors to inconsistent standards.\footnote{For instance, one co-author encountered a reviewer who argued for a paper to be rejected because it used an OpenAI model.} %
As we will show, acceptance decisions can also hinge upon accusations of whether the system goes beyond putting a ``wrapper'' around LLMs%
---a determination that is not well-defined. %
Thus, we need to better understand the situation of authors and reviewers to set realistic, balanced expectations, as more systems will continue to incorporate LLMs and grow in complexity with agents and tool use.

In this work, we report an interview study of 18 HCI scholars' experiences and challenges when building, reporting, and reviewing LLM-integrated systems, %
and position these experiences in the broader context of past and current changes in HCI research. 
Our initial goal was to provide a set of considerations for the community on reporting these systems, in a manner that balances the practical realities of authors' time, cost, and page limitations, scientific concerns such as reproducibility and transparency, and the unique norms and expectations of HCI research. 

Overall, we find evidence that \textit{the inherent uncertainty of LLMs has eroded traditional trust-building norms between authors and reviewers}, resulting in reviewers at large demanding more details to validate claims, such as increasingly expecting technical evaluations outside of user studies. Authors perceive reviewers at HCI venues such as CHI and UIST exhibit wide variability in demands for this paper type, above and beyond the baseline fluctuations %
of peer review.
We also describe how authors approach designing and reporting LLM-integrated systems in HCI research, from how LLMs seem to shift HCI towards more technology-centric design~\cite{yang2020re}, to how authors balance the pressure to report more details under page and word limits by shifting details into appendices and code repositories. Finally, many tensions appear to derive from a broader clash of values between HCI and more technical computing fields such as ML and NLP, especially on what constitutes a valid contribution and an appropriate level of technical rigor. 

From our study findings, we synthesized a set of considerations and suggestions for authors, reviewers, and the HCI community around papers that report LLM-integrated systems (Section~\ref{recommendations}). We iterated on this section through additional feedback from six (6) HCI researchers with 15+ years experience each. To begin, we outline relevant meta-research and historical context, followed by a description of our interview methodology, before proceeding to the findings, recommendations, and discussion. %

\indent {\bf Positionality.}
We are one Full Professor (Syriani, approximately 20 years of experience), one Assistant Professor (Arawjo, about 10 years of experience), and one graduate student (Navarro, formerly a full-time academic with about 15 years of experience) %
with a combined experience of nearly five decades in publishing at HCI and software engineering venues. The two professors have been developing and publishing LLM-integrated systems since 2023. 
We have seen issues emerge around these paper types during the peer review process for papers and grants, whether as authors, reviewers, meta-reviewers, and in program committee (PC) meetings. In particular, Arawjo served as a PC member for UIST '24, '25, and CHI '26; as a meta-reviewer, 2AC, or external reviewer, they have participated in reviewing around 50 LLM-integrated system papers. They also wrote a widely-circulated blog post about these experiences~\cite{arawjollmwrapper}, which informed the new desk rejection guidelines of UIST '25.  %
Additionally, we have conducted work in the prompt engineering and LLM evaluation space~\cite{arawjo2024chainforge, shankar2024validates, zhang2025chainbuddy, syriani2024screeningarticles, syriani2022recommending, syriani2024toward, semantic_commit}, with Arawjo teaching courses on building LLM-integrated systems. These experiences informed our motivation, questioning, and considerations. %
Finally, to guard against bias, we iterated on our considerations section with the feedback of six expert HCI researchers, each with more than 15 years of experience, which we elaborate on in \S\ref{recommendations}. %

\section{Context and Related Work}

\begin{figure*}[t]
  \centering
  \includegraphics[width=\linewidth]{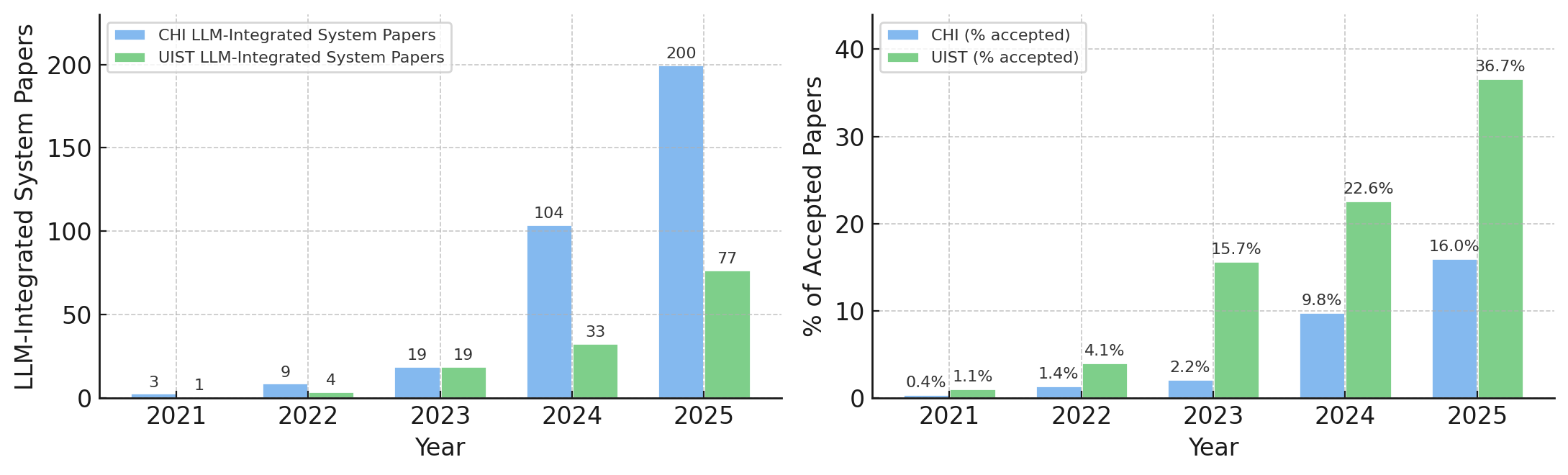}
  \caption{The number of papers with LLM-integrated systems published at CHI and UIST since 2021, both as absolute counts and the percentage share of all accepted papers.}
  \Description{Bar charts showing the number of LLM-integrated system papers published at CHI and UIST from 2021 to 2025, as well as the percentage proportion of such papers compared to the full accepted count. Counts rise sharply and consistently over time. Both conferences had very few papers in 2021 and 2022. By 2023, counts reached about 20 at each venue. In 2024, CHI grew to over 100 while UIST reached about 30. In 2025, CHI more than doubled to just over 200. By UIST 2025, around 36.7\% of all papers reported LLM-integrated systems.}
  \label{fig:chi-uist-counts}
\end{figure*}

This paper focuses on \textit{LLM-integrated systems} reported in HCI papers, defined here as systems that query an LLM at any stage of program execution. This paper type corresponds to Pang et al.'s ``LLMs as system engines''~%
\cite{pangLLMificationCHI2025}. %
Other categories of work---such as ``LLMs as objects of study'' which examine training datasets or biases inherent to LLMs---fall outside our scope.

Our work relates to emerging meta-research around the impact of AI on academic publishing processes and research more broadly. Across HCI, conferences are experiencing an influx in submissions: UIST 2025 experienced a 56\% increase in 2025, %
on top of 26\% the previous year, %
while CHI has experienced a 58\% increase since 2023 \cite{tanaka2025assistedDeskReject}. The Technical and Paper Chairs of CHI 2026 remarked that the influx of papers has caused a ``diluted review process'' and ``collapse in reviewer recruitment'' that motivates a new assisted desk reject policy~\cite{tanaka2025assistedDeskReject}. CSCW has also faced %
``sharply increased'' submissions, with organizers warning that without changes, ``our peer review process will not be sustainable'' \cite{cscw2026_reviewprocess}. 

Our accounting of LLM-integrated system papers at CHI and UIST from 2021 onwards (Figure~\ref{fig:chi-uist-counts}) indicates that the number of papers increases each year, and roughly doubled from CHI 2024 to CHI 2025.\footnote{We produced this chart through search of the ACM Digital Library with keywords ``LLM'' or ``language model'' appearing together with ``system'' or ``interface'' anywhere in the paper, and checking the paper PDF that 1) authors describe a system they built and 2) the system queries a large language model at some point in its execution. We manually visited each page and scraped paper metadata with a custom Chrome plugin.} The doubling is not fully accounted for by the 17.8\% proportional increase in accepted papers alone. At UIST 2025, a little over one in three papers report LLM-integrated systems. The general number of papers that reference LLMs is also increasing: at CHI 2025, Pang et al.~\cite{pangLLMificationCHI2025} presented a literature review of all papers that mention LLMs at CHI up to 2024, providing a topical overview. They find five roles that LLMs play in research---as system engines, as research tools, as participants and users, as objects of study, and as perceived by users. They point out that LLMs are most prominently used for writing, and raise concerns over the  over-reliance on proprietary models like OpenAI and the absence of reporting prompts in some papers, but cannot explain why these choices are made. %

There are also indications that paper submissions with LLM-integrated systems have especially caused problems for peer review. For instance, in an opinion piece to the VIS BELIV workshop titled ``We Don't Know How to Assess LLM Contributions in VIS/HCI,'' Crisan reports personal experiences across four conference program committees on common reviewer critiques of LLM-based papers, such as asking to evaluate against multiple LLMs, pre-print culture disrupting peer review, and general fatigue with LLMs, and suggests solutions such as asking authors to justify usage, evaluate prompt sensitivity, and desk reject more papers~\cite{crisan2024we}. In 2024, Arawjo also made a blog post titled ``LLM Wrapper Papers Are Hurting HCI Research''~\cite{arawjollmwrapper}, which circulated in the community and has been cited in other work, including Crisan. This post hypothesizes that papers from ML and NLP conferences that can be summarized as ``LLMs applied to X problem'' are flooding HCI conferences, many of which engage little with past HCI research. Both UIST 2025 and CHI 2026 guidelines added new desk-reject policies which reflect this point. %
These two pieces suggest that the topic is deserving of  attention, yet remain opinion pieces, with limited evidence presented to provide nuance to the situation and ground claims. %

HCI research, especially in systems, has had a long history of debates around peer review. %
At UIST 2007 and CHI 2008, Olsen and Greenberg \& Buxton, respectively, raised concerns that usability studies, especially controlled ones, were becoming expected methodologies by reviewers to the detriment of other HCI approaches and contributions~\cite{olsen2007usability, greenberg2008usability}. %
In 2009, Landay %
felt the situation for systems research had ``gotten worse,'' with reviewers ``asking you to do 2-4 years more work'' by re-implementing proprietary systems just to conduct a controlled study, and asking for large numbers of study participants to justify scientific value~\cite{landay2009_giveup_on_chiuist}. Recently, Soden et al. raised similar alarm bells in qualitative HCI research, reporting that ``increasingly all qualitative research... is being evaluated from the perspective of positivism'' \cite{soden2024evaluating}. Examples include reviewers narrowly focused on quantification, generalizability, and reproducibility, such as calling for demographics, code counts, inter-reliability scores, and more participants, regardless of research goals. Crabtree argues that this ``problem of methodologically incongruent (positivistic) reviews of qualitative research... deprive HCI of crucial insights into the relationship between technology and society,'' pointing out that positivism is incompatible with interpretivism \cite[p. 2]{crabtree2025h}.\footnote{A full account of the differences is provided by Crabtree~\cite{crabtree2025h}, but we provide a brief overview. Positivism ``seeks to explain human conduct in causal terms,'' emphasizing reproducibility, controlled experiments, quantification, and generalizability, and is what many people might traditionally call "science"~\cite{crabtree2025h}. Interpretivism, by contrast, emphasizes collecting rich qualitative data and the role of society and culture on how humans construct meaning and interpret experiences. In HCI, design methods arguably form a third perspective---see, e.g., research through design~\cite{zimmerman2007research}.}  %
These concerns are relevant for systems research, which often includes qualitative methods and design contributions. %
Taken together, they suggest that reviewers continue to expect and apply positivist notions of science to HCI research. %
More broadly, these problems may reflect how HCI as a field has historically felt its precarious position in computing; for instance, %
Jonathan Grudin remarked: ``CHI is very insecure, and feels that if we accept rough but interesting and important work, we will be seen as not rigorous and lose the respect of Computer Science colleagues''~\cite{landay2009_giveup_on_chiuist}. In brief, longstanding clashes of value and methodology for scientific rigor ---between interpretive research in HCI and traditional, positivist concepts of scientific rigor in more technical computing fields---provide important context for situating our findings, and we shall return to them in Discussion.%

Finally, our work contributes to broader history of frameworks for reporting AI systems. In the past, reporting challenges in adjacent fields to HCI, such as machine learning, have led to frameworks and attempts at standardization. For instance, %
ML frameworks like model cards~\cite{mitchell2019model}, datasheets for datasets~\cite{gebru2021datasheets}, and HELM~\cite{liang2022holistic} focus on model-centric details, such as provenance, evaluation data, and authorial intent. %
A recent pre-print in software engineering focuses on rigor in technical evaluation such as requiring comparison to open models~\cite{baltes2025evaluation}. However, how suitable these requirements are for an HCI context remains unclear. %

\section{Methodology}

To better understand authors' and reviewers' experiences when preparing and publishing papers that report LLM-integrated systems in HCI, we conducted an interview study with authors of such papers. We focused on this population because, under \textit{quid pro quo} reviewing expectations at venues such as CHI, these authors are also likely to have served as reviewers.

\subsection{Data collection}
We initially recruited 17 participants, supplemented with an email response from one professor (P18) who could not attend the official interview but granted us permission to include their input. While preparing our findings, we also recruited one additional participant (P19), a senior industry researcher and long-standing member of the HCI systems community, to provide a broader historical context on peer review and further validate our findings.\footnote{This session was unique. We first ran the standard interview, whereupon P19 independently reflected several themes, including the evolving perceptions of LLMs, the increased uncertainty in reviewing expectations for these papers, the clash of communities, and using ``wrapper'' as a metric for rejection. After this interview, we directly relayed some of our early findings to P19 to get their feedback and perceptions.} Recruitment strategies included cold emails to first authors of relevant CHI and UIST papers, social networks, Slack message boards, and word-of-mouth. In total, we invited 
64 people. Interviews lasted approximately one hour and were conducted over video conferencing. Two coauthors served as interviewers—one graduate student and one professor—each conducting roughly half of the sessions. As an incentive, participants were offered entry into a lottery for \$40 CAD, with a win rate of about 30\%.
In addition, we obtained five reviews and two rebuttals from a survey.\footnote{ACM's official policy allows the sharing of anonymized reviews that authors \textit{received} for their papers~\cite{acm-peer-review-faq}. Through email with the ACM Director of Publications, we obtained additional clarification that sharing rebuttals was permissible. We stressed that participants should not share reviews they had written as reviewers for other papers.}
Participants were hesitant to share these materials, often seeking co-author or advisor approval first. More frequently, they discussed the content informally during interviews, where reviews and rebuttals served as prompts to aid recall. To further triangulate insights, we included four sets of reviews from our own past submissions of LLM-integrated system papers to HCI venues (all unrelated to this project). Our study received ethics board approval.

\subsection{Participant demographics and experience}
Our sample included six participants in professorial or postdoctoral positions, one industry researcher, and the remainder primarily PhD students in the latter stages of their PhD (1 industry researcher, 4 professors, 2 postdocs, 8 PhD students, 2 master's students, and 1 undergraduate; 4 female, 14 male). Participants reported being active in the field for a median of 6 years (range 2--25, average 7.2). They served as lead or advising author on a median of 3.5 LLM-integrated system papers, and in a secondary role on 1.5 more. For papers of this type, they reported a median of 3 accepted, 1.5 rejected, and 2 in-progress or in-submission articles. As reviewers, participants had evaluated a median of 10 LLM-integrated system papers in HCI venues, with five more senior researchers estimating upwards of 30–40 each.
Participants had published across a range of venues: CHI (13 participants mentioned), UIST (11), IUI (5), with additional mentions of DIS, CSCW, and ASSETS. Five participants also had publishing experience outside HCI, in venues such as EMNLP, NeurIPS, VLDB, SIGMOD, and ICSE, enabling comparative reflections across research communities. To protect our participants' identities within the HCI subset working on LLM-integrated systems, we do not provide a demographic table.

\subsection{Research objectives and interview protocol} 
Our goal was to understand participants' experiences in conducting, publishing, and reviewing LLM-integrated systems. The semi-structured interview protocol was organized into three parts: 1) %
participants’ experiences as \textit{authors} building systems, reporting them in papers, and navigating the review process; 2) %
their experiences as \textit{reviewers} of others' LLM-integrated papers and their broader perceptions of how such work is received within HCI; and 3) %
their guidelines, advice, or standards for future reporting.

Throughout, we emphasized what participants found \textit{unique} about working with LLM-integrated systems compared to systems without LLM components. When relevant, participants were prompted to view their authored papers and received reviews to aid recall. Following grounded theory and semi-structured interviewing methods~\cite{charmaz2006constructing}, we refined the protocol iteratively, adding questions on topics such as open model usage, reviewer expectations for non-LLM papers, the role of reproducibility, and whether AI/LLM terms should appear in titles or abstracts.  

To mitigate bias, we refrained from introducing terminology unless the participant used it themselves, such as the terms ``wrapper'' and ``uncertainty.'' We phrase questions  to allow disagreement; e.g., ``Did you encounter any particular or unique challenges when designing LLM-integrated systems compared to non-LLM systems?''

\subsection{\bf Data analysis}
We conducted inductive thematic analysis. Two coauthors independently performed line-by-line open coding of interview transcripts in a shared spreadsheet, generating 1,458 initial codes. Consistent with grounded theory~\cite{charmaz2006constructing}, our initial codes erred on the side of specificity and summary, with the intention of clustering codes later through focused and axial coding. We manually clustered all codes in an affinity diagram in canvas software to identify emergent themes and connections across themes. All coauthors collaboratively reviewed the themes to refine gaps, strengthen coherence, and clarify theme naming. Although the review and rebuttal sets we collected were limited, they were used as primary source exemplars to illustrate themes identified from the interviews. In preparing findings, we gave particular attention to the perspectives of researchers with 10+ years in the field, who have extensive reviewing experience and influence junior researchers.

\subsection{\bf Limitations}
Our recruitment may be subject to selection bias, as authors with strong opinions---particularly negative reviewing experiences---may have been more motivated to participate. Also, we may have attracted participants who are more interested in discussing with us, familiar with our work, and active online.
Recruitment also skewed toward younger researchers: despite direct outreach, professors were often unavailable or redirected us to their students. While we recruited P19 to help validate our findings and hedge against this limitation, perspectives from other more senior scholars may still be underrepresented.  Our sample also centered primarily on academia, with only one current industry researcher.

Another limitation lies in our reliance on participant recall. While reviews and rebuttals provided some primary material, sharing such documents conflicts with longstanding norms of closed review in HCI, making them difficult to obtain. %
To empathize with authors' perspectives, we also focused primarily on reviewers who had authored LLM-integrated system papers; perspectives of reviewers who have not authored such work may differ.

We therefore position our study as an initial step toward surfacing hidden tensions within the community, rather than a definitive account. We hope our findings, analysis, and considerations stimulate broader community discussion.

\section{Findings}

Our analysis revealed that many challenges faced by participants---whether as authors or reviewers---centered around the problem of \textit{establishing trust in the face of LLM uncertainty}. Specifically:
\begin{enumerate}
    \item For authors, trust meant gaining confidence that their systems would perform reliably for target users, despite uncertainty in LLM outputs. 
    \item For reviewers, trust meant gaining confidence that authors' claims were credible and that systems would perform as described, despite uncertainty in both LLM behavior and the reliability of authors' reporting. 
\end{enumerate}

While these are not new problems in HCI systems research, participants emphasized that they are uniquely amplified by the intrinsic uncertainty of LLMs as stochastic, black-boxed models, as well as by the hyperbolic public discourse around AI. Before the recent wave of LLMs, most system components were predictable, with bounded inputs and outputs. The main source of uncertainty stemmed from variance in user behavior, which authors typically addressed through user studies and experiment design. %
As a result, system papers often relied on screenshots, examples, and descriptive accounts of technical components, without the need for extensive technical evaluations prior to user studies.

With LLM components, uncertainty arises from the models themselves. Participants noted that this uncertainty, combined with heightened skepticism due to AI hype and placebo effects~\cite{ai_placebo_effects,ai_nocebo_effects}, makes a positive user study insufficient on its own to establish credibility. Reviewers, in particular, were described as seeking stronger methodological justifications and more extensive reporting than they would expect for non-LLM system papers. In response, authors reported adopting additional strategies such as conducting technical evaluations, explicitly justifying the inclusion of LLMs, and appending extensive technical details.

Table~\ref{tab:uncertainty} summarizes the main sources of uncertainty raised by participants in authorship and reviewing roles, along with strategies used to mitigate them. Some of these are not unique to LLMs, but are amplified by the uncertainty of LLM outputs and capabilities. We emphasize that these strategies are not necessarily \textit{standards} that authors/reviewers should be held to, but rather interviewees' current learned strategies to re-build trust in the face of uncertainty.
In the following subsections, we unpack how these strategies manifest across the stages of building, reporting, and reviewing LLM-integrated systems, and how they collectively serve to re-establish trust in the face of LLM uncertainty.\footnote{Because interviewees could draw on both authorship and reviewing experiences, we refer to ``authors'' and ``reviewers'' directly for brevity, rather than specifying ``in their role as an author or reviewer.''}
\begin{table*}[h]
\centering
\small
\renewcommand{\arraystretch}{1.2}
\begin{tabular}{p{0.3\linewidth}p{0.65\linewidth}}
\toprule
\textbf{Uncertainty about...} & \textbf{Strategies to build trust} \\
\midrule
\multicolumn{2}{l}{\textbf{Role:} \textit{Builder}} \\
Whether LLMs can accomplish the task & Rapid ad hoc testing of prompts and models, trying varied prompting strategies or pipelines early on. \\
What model to choose & Compare models in initial trials, considering latency, cost, resource access, UX goals, reproducibility, and ethics. \\
How to prompt effectively & Experiment with multiple prompting strategies rather than fixating on one. \\
How to evaluate performance & Iteratively develop evaluation rubrics; ask team members or domain stakeholders to annotate outputs. \\
Context to include and how & Test different amounts and types of context to compare performance. \\
Robustness of LLM component behavior & Create datasets of diverse inputs representative of the task domain and evaluate LLM outputs. \\
Real-world use (e.g., open user prompts) & Conduct human validation studies with domain experts or target users before launching user studies. \\
\midrule
\multicolumn{2}{l}{\textbf{Role:} \textit{Author}} \\
How much details to report around the LLM component & Report only details critical for understanding/evaluating contributions; use appendices, supplementary materials, or code releases to extend transparency. \\
Reviewer expectations around types of evaluation required & Add technical evaluations or pre-study validations to satisfy technical reviewers; provide extensive details and justifications to address skeptical reviewers. \\
Reviewers' idiosyncratic positions regarding the use of LLMs & Explicitly discuss ethical risks and societal ramifications, especially in sensitive domains; choose topics and framings carefully; encourage venues to set clearer guidelines for LLM paper evaluation. \\
\midrule
\multicolumn{2}{l}{\textbf{Role:} \textit{Reviewer}} \\
Author's level of rigor in engineering the system & Authors can informally report their prompt/LLM component engineering process, noting alternative models or strategies tested to contextualize design decisions. \\ 
Why authors chose to use LLMs & Authors can justify LLM use in relation to design goals, and de-emphasize LLM/AI in the title/abstract/intro if it is not the primary contribution. \\
Failure modes of LLM components & Authors can explicitly describe LLM limitations and include illustrative examples of system errors. \\
How the LLM components integrate inside the system & Provide an end-to-end system workflow diagram with truncated input/output examples; report prompts central to claimed contributions. \\ %
Representativeness of authors' examples of system behavior & Authors can include examples derived from user studies and/or small-scale technical evaluations across representative inputs. \\
Robustness and generalization of system & Authors can include technical evaluations across representative datasets. \\
What parts of technical architecture contributed to  positive performance & Authors can run ablation studies to identify which components contribute to performance, if claiming a technical contribution. \\
Whether participants' positive responses in a usability study can be trusted %
& Authors can compare against AI-powered baselines to address potential AI placebo effects\cite{ai_placebo_effects}; include technical or field evaluations beyond usability studies. \\
Whether technical approach will remain relevant after future AI advancements & Authors can frame contributions around interaction design or insights into user behavior. %
\\
\bottomrule
\end{tabular}
\caption{Sources of uncertainty for authors and reviewers, and strategies to resolve/mitigate them that were mentioned by participants in both author and reviewer roles.}
\Description{Table with two columns, showing "Uncertainty about..." in left column, and "Strategies to build trust" in right column. Additionally, there are three overall sections for "builders", "authors", and "reviewers." Listings reflect findings in the paper.}
\label{tab:uncertainty}
\end{table*}

\subsection{Authors designing LLM-integrated systems}

\begin{quote}
    \emph{``The target is moving, and like your bow is  completely f***ed, and you also only have 5 arrows.''} --P16
\end{quote}

\subsubsection{\bf Technology in search of a problem: LLM capabilities can bias problem selection and design} \label{4-1-1} 
Participants described starting projects by probing LLM capabilities with ad hoc prompting, then pursuing problems where the model 
seemed competent and discarding others. P10 contrasted this with their lab's pre-LLM workflow, which began with user needs, low-fidelity 
prototyping, and early feedback. Now their students often start with a broad idea, try it in ChatGPT or Claude, and then pick the domain and task where the model looks strongest (\textit{``we're literally just trying different things and doing an eyeball test,''} P10).

This ``technology-driven'' development process resembles Yang et al.'s revised UX design framework of a ``technology-driven design innovation process,'' in which designers first ``characterize tech capabilities'' \textit{prior} to ``understanding user activities''~\cite[p. 3]{yang2020re}. This shift was not due to ignorance of user-centered design, but a strategy to avoid investing in infeasible directions. Still, participants worried that letting proprietary LLMs steer feasibility testing risks biasing HCI systems research toward problems that models can already solve, rather than those most grounded in user needs or novel interaction paradigms.

\subsubsection{\bf Open-source vs. Proprietary: Model selection mediated by cost, latency, and performance trade-offs.} \label{4-1-2}
Participants acknowledged ethical and reproducibility concerns with closed-source APIs~\cite{pangLLMificationCHI2025}, which companies can deprecate without warning~\cite{ma2024schrodinger}. Yet, most preferred proprietary models. As justification, they cited easier prototyping, access to the latest capabilities (e.g., multimodality), reduced development time due to better performance, and closer alignment with baselines familiar to users and reviewers (e.g., if comparing a system to OpenAI's ChatGPT, choosing a open model like Qwen could introduce a confounding variable to experiment design). Grants or free credits also lowered costs.  

Open-source options were attempted but often abandoned due to poor performance, higher resource demands, and especially latency---critical in interaction studies where delays shape user perceptions and study outcomes. P8 described switching from an open model after spending excessive time decomposing tasks, prompt-engineering, and adding guardrails.
While some worried about reproducibility, many felt this mattered less in HCI than in ML/NLP, since HCI emphasizes prototyping and proof-of-concept over strict benchmarking. 
At the same time, several participants chose open models because they were designing systems in contexts where user privacy is central. However, open models were not universally perceived as ethical; for instance, one participant felt that proprietary models are safer when handling unexpected inputs. Model choices were framed as problem- and domain-specific decisions rather than universal imperatives.

\subsubsection{\bf Uncertainty of LLMs distinguishes them from traditional code-based systems} \label{4-1-3}

When asked what distinguishes LLM-integrated systems from traditional code-based systems, many participants highlighted the non-determinism of LLMs as the unique factor. Five explicitly used the term ``uncertainty'' to describe this difference (P1, P2, P13, P15, P16). This uncertainty seemed to erode their confidence in system effectiveness and compounded the challenge of anticipating user behavior: because LLMs can accept virtually any input, participants noted difficulty in predicting how users would interact with their systems and how the models might respond in practice.

\subsubsection{\bf Authors' strategies to build confidence in their systems in the face of LLM uncertainty.} \label{4-1-4} The process of LLM component engineering appeared necessarily iterative, largely unsystematic, and  stable across participants. %
Typically, authors began by experimenting with multiple models and prompting strategies in an ad hoc manner to test feasibility. Once a model and initial strategy were chosen, they refined prompts through four common practices: (1) eyeballing outputs and enumerating failure modes individually or in teams; (2) constructing small, custom datasets over representative inputs to probe variability and iterate with greater confidence; (3) sharing outputs with team members, stakeholders, or domain experts for additional feedback before formal user studies; and (4) conducting scaled-down pilot deployments. P12, for instance, created a shared Google Docs for prompt versions and example outputs, shared with stakeholders to provide feedback. Some authors were reluctant to report these processes in papers (and even discuss them with us), expressing concern that the lack of systematic rigor, especially around prompt engineering, might undermine their credibility. %
In some cases, technical evaluations were retroactively added after user studies concluded, often to address reviewer requests. Participants could also omit known failure modes of their system and methodological details: P3, for instance, recalled how a small change to a prompt disrupted an in-the-wild workshop, but noted this detail was not reported in their paper. %

\subsubsection{\bf Stop when it is ``good enough'' for a user study} \label{4-1-5}
Seven participants reported stopping system iteration once they deemed the system ``good enough'' to run a user study. They justified this choice by referencing established HCI research values, which prioritize enabling empirical investigation through functional prototypes over pursuing fully optimized system performance. For example, P6 felt it was not necessary to ``optimize'' since their goal was typically to showcase interactions or probe user behavior. %

\subsection{Deciding What and How Much to Report}

\subsubsection{\bf Reporting prompts should be selective, focusing on what is critical to validating claimed contributions, difficult to engineer, or transferable} \label{4-2-1}
Contrary to blanket recommendations to ``report all prompts''~\cite{pangLLMificationCHI2025}, participants largely agreed that reporting depends on three factors: (1) how \textit{critical} the prompt or system detail is to the paper's contribution or claims, (2) how \textit{difficult} it was to engineer the LLM component, and (3) how \textit{interesting} the detail may be for readers, especially if it can be adapted in other contexts. As LLMs become more deeply embedded in software---such as ``agent'' systems with dozens or even hundreds of prompts---documenting every prompt in the paper was seen as intractable, and potentially distracting from contributions.  

For example, if a technical contribution hinges on an LLM (e.g., a visualization technique that requires specific LLM-driven functionality), then reporting those prompts is essential. In contrast, for a peripheral feature (e.g., a ``summarize text'' button implemented with the first working prompt authors tried), participants felt it was sufficient to disclose that it was simple to engineer, without detailing the exact wording. In such cases, requiring exhaustive prompt reporting and evaluation would be disproportionate. As LLMs become routine building blocks, participants expect that selective reporting, focusing on prompts central to claims, challenging to engineer, or broadly useful to others, will become the norm.

\subsubsection{\bf Authors use appendices, supplementary material, and repositories to offload LLM details} \label{4-2-2}
When asked where they documented prompts, nearly all participants said they placed them in the appendix. This strategy allowed authors to comply with calls for transparency while avoiding clutter in the main text. However, tensions arose as systems grew more complex and prompts became longer: including dozens of prompts verbatim was seen as infeasible.  

To address this, some authors shifted details into supplementary material or open-source repositories, framing this as a more practical and scalable way to provide access. Others extended this strategy to architectural details, which were often relegated to appendices rather than the main body of the paper. While this practice helped manage page limits and unwanted distraction from core claims, it also raised questions about how much reviewers actually read or verify supplementary content. 

\subsubsection{\bf Use high-level sketches of system architecture to show how data flows through LLM components}  \label{4-2-3} 
As both authors and reviewers, participants emphasized the value of reporting a ``sketch'' of the overall architecture rather than exhaustive  dumps of prompts. They recommended using diagrams, tables, or abbreviated examples of prompt templates to illustrate how data flows from input(s) through LLM components, paired with representative input/output examples. This kind of high-level overview was seen as more informative and accessible, especially as systems become more complex and may involve dozens or even hundreds of prompts. %

\subsubsection{\bf Learned strategy: De-emphasize LLMs in titles and abstracts to foreground non-LLM contributions}   \label{4-2-4}
An unexpected theme centered on paper framing. Authors described a dilemma: on the one hand, they felt that mentioning ``LLM'' or ``AI'' in a title, abstract, or introduction section could boost visibility, attract readers, and even create career opportunities for graduate students. On the other hand, authors worried that foregrounding LLM usage might trigger dismissive reviews or cause reviewers to focus their evaluation on LLM components and overlook other contributions.  
To navigate this, five participants reported deliberately de-emphasizing LLMs in their framing. For example, P1 omitted any mention of LLMs until their implementation section, after a team discussion. %

\subsubsection{\bf Authors valued reproducibility, but emphasized \textit{interaction} over exact performance}   \label{4-2-5}
Participants distinguished between reproducibility as a scientific ideal and how they approached it in practice. In HCI venues, many described striving for what might be called \textit{soft reproducibility}: ensuring that an interaction technique could be recreated with another model and comparable prompt engineering effort, rather than replicating exact outputs or performance. For instance, P14 downplayed the issue of model deprecation~\cite{ma2024schrodinger} for reproducibility, anticipating that future LLMs would converge toward similar or improved capabilities.

One participant with experience in NLP venues contrasted this with \textit{hard reproducibility}, which they characterized as critical in NLP because progress often resembles ``hill climbing,'' requiring precise replication of models, parameters, and prompts. They argued that HCI research is more about creating new ``hills'' (novel interaction paradigms) where reproducibility of the user-facing interaction, rather than the exact technical configuration, is paramount. As further evidence of this disciplinary distinction, they noted that, unlike many computing fields, HCI venues routinely accept system papers without requiring code submission.  

Still, participants expressed that \textit{context} matters: in sensitive domains such as health, full transparency and hard reproducibility were considered essential, whereas lower-stakes domains warranted less detail. However, most participants supported open-sourcing code where possible to strengthen reproducibility, reduce the burden of lengthy appendices, clarify system architecture, and enable others to compare to or build upon prior work.

\subsubsection{\bf Learning reporting practices by observing published work}   \label{4-2-6}
A few participants described looking to prior publications, whether in HCI and others, for inspiration on how to report LLM components within a limited space. For example, P3 pointed to a multi-agent paper that summarized components in a table with columns for the  goal of the component, model used, overarching prompt, and truncated input/output examples; P4 adopted a similar approach. P15 looked to ML papers for guidance on how to present technical evaluations.  
When asked about potential guidelines, participants also emphasized the value of referencing exemplar papers, which they considered as practical resources for learning and disseminating reporting practices.

\subsection{Experiences in Peer Review and as Reviewers, or: ``Is it an LLM Wrapper?{}''}

\begin{quote}
    \emph{``It is not the business of a random reviewer to tell an institution in South Korea [what to do]... There are of course people who don't subjectively like these thin LLM wrappers... [but] I think you should have better reasons to reject the paper than `I don't like this type of work.' You should be able to find the scientific reasons to reject something.''} --P11
\end{quote}

\subsubsection{\bf Authors perceived reviewers as unusually harsh and inconsistent toward LLM-integrated systems}  \label{4-3-1}
While participants acknowledged broader quality issues with HCI peer review, most felt that papers involving LLMs attract especially severe and inconsistent scrutiny. They described dismissive reactions to the mere presence of an LLM component, reviewers imposing strong personal or moral stances on authors, and %
HCI contributions without technical innovations being undervalued when LLMs are employed to drive prototypes. As P3, P5, and P6 put it, some reviewers have immediate ``knee-jerk'' reactions once they see an LLM. One participant recounted presenting a poster at CSCW where ``two to three attendees walked away'' upon being told the system used an LLM.  

Some authors attributed this response to reviewers' political or ethical concerns. P11, for example, argued that Western reviewers had become ``moral guardians'' policing global research practices, and stressed that ethics board approvals should serve as the benchmark rather than reviewers' personal values. Similarly, P13 characterized reviewers as ``gatekeepers'' treating LLM-based systems with unusual suspicion compared to other technical choices. P5 worried that such gatekeeping discouraged exploration of LLMs in sensitive domains, such as health, limiting HCI's potential impact.  

A few participants noted that some reviewers appeared overly positive about papers simply because they used LLMs. Rather than a counterpoint, this was presented as further evidence of wide variability in reviewer expectations. Interestingly, P19 speculated that this variability may be exacerbated by the decline of in-person program committee meetings, which had long served to coordinate and ground community values.

\subsubsection{\bf Justifying technical innovation is harder: solving a problem is `no longer interesting enough'}  \label{4-3-2}
Participants perceived a shift in reviewing standards for LLM-based system papers. Around 2023, they felt that the mere inclusion of an LLM could make a paper publishable. Now, with the \textit{``low-hanging fruit''} gone (P6) and many LLM-integrated systems already published, reviewers appear to expect substantially more. Authors also worried that popular domains have become saturated, making it harder to establish novelty. For example, in one review we collected, a reviewer dismissed the contribution of a writing system as lacking novelty given the \textit{``crowded space''} of LLM-based writing tools.  

Several contrasted this to pre-LLM norms, where building a system itself often required years of painstaking work~\cite{landay2009_giveup_on_chiuist}, and the act of solving a problem and running a user study could be sufficient. In contrast, participants felt that, because LLMs can now solve many problems \textit{``out of the box,''} reviewers view problem-solving alone as uninteresting (\textit{``we just solved yet another problem with an LLM,''} P6).  
In response, three strategies emerged for justifying innovation under these new expectations:  
\begin{itemize}
    \item \textbf{Novelty as insights into user behavior:} framing the system primarily as a probe to surface new understandings of user practices and behaviors.  
    \item \textbf{Novelty as interaction design:} exploring interaction designs beyond natural language (e.g., moving away from chatbots and prompting to explore novel UI-level designs).  
    \item \textbf{Novelty as problem domain:} selecting under-explored domains where LLMs have not yet been widely applied.  
\end{itemize}

\subsubsection{\bf Conveying ``effort'' in engineering or design mediates paper acceptance} \label{4-3-3}  
As a strategy to justify innovation, seven participants felt that signaling authors' effort, thoughtfulness, or system complexity could positively influence paper acceptance. For example, P12 emphasized the importance of reviewers recognizing the ``thought'' and uniqueness invested in system design, while P2 described highlighting the sophistication of their front-end UI and the difficulty of prompt engineering as evidence of technical novelty.  

At the same time, participants noted a risk that emphasizing complexity could undermine credibility. For P16, encountering a highly complex system description section created uncertainty about whether the complexity was truly necessary. In such cases, reviewers may suspect that authors are presenting complexity not out of research need, but as a strategy to strengthen acceptance.

\subsubsection{\bf Technical evaluations of LLM components are becoming expected, outside of a user study.} \label{4-3-5}
Fourteen participants perceived that LLM-integrated system papers must include a technical evaluation (whether automated metrics or expert raters) %
\textit{in addition to} a user study. For some, adding a technical evaluation was a learned strategy to avoid reviewer push-back, after experiencing rejections or resubmits that hinged on the absence of one. %
Participants felt this marked a departure from pre-LLM HCI systems work, where user studies alone, often with small sample sizes, were typically seen as sufficient. With the uncertainty of LLM behavior, reviewers expressed skepticism that a positive user study demonstrates robustness, citing concerns about authors cherry-picking positive behavior (\textit{``...it could be representative, but... maybe a successful example... is like the only one [working] example out of a hundred''}, P1), positive user bias in studies, demand characteristics, and AI placebo effects~\cite{ai_placebo_effects, ai_nocebo_effects, IARYGINA2025103379}. The open-ended nature of LLM interactions further compounds doubts about generalizability.  

While some participants acknowledged that technical evaluations can productively complement user studies, others worried that they are becoming a default demand with high expectations for rigor, irrespective of the authors' stated contributions. P9, reflecting on a rejection, argued: \textit{``I don't think [reviewers] should have asked for a [technical] evaluation. I think they should have asked for a better rationale... It doesn't make sense to do a very very rigorous analysis when you're already 95\% confident it's useful. Why not just do a better job at the user study?''} P10 similarly emphasized that evaluation requirements should align with authors' claims, yet described reviewers who demanded technical evaluations regardless. P15 felt such reviewers missed the ``point,'' focusing narrowly on reliability while overlooking interaction design insights. P3 described reviewers of a rejected paper who suggested fine-tuning a model to enhance the ``technical contribution,'' a demand they felt was immaterial to their core claims.

Reflecting on this shift, P19 confirmed that technical evaluations had not been a standard expectation in software systems research before, but noted they had long been required in \textit{ML}-integrated HCI work to validate model performance during training.

\subsubsection{\bf Reviewers demand more extensive details and context than for non-LLM systems} \label{4-3-6} 
In addition to technical evaluations, reviewers were seen as demanding substantially more details and contextual information from LLM-integrated system papers than from traditional systems. Twelve participants described either receiving such requests or making them themselves as reviewers. P4 summarized this shift: \textit{``LLM system papers often warrant much more additional context from the author's side to the reviewers to sufficiently convince the reviewers of the work. And that is trickier to do, especially when page limits are not fixed.''}

\subsubsection{\bf Clash of community values: Tensions between HCI and reviewers from more technical fields such as NLP and ML} \label{4-3-7}
Ten participants framed tensions in peer review as part of a broader clash of communities between HCI and %
more technical computing fields, especially ML and NLP. For instance, P15 felt their reviewers expected standard measures from NLP to validate LLM components, and recruited expert raters to comply. They perceived that the two communities were in the process of merging, analogizing the situation to the emergence of information visualization, which grew out of a collision between computer graphics and HCI. While acknowledging potential benefits of cross-pollination, they questioned whether ML/NLP methods were appropriate for HCI research: \textit{``Do the evaluations that we have in the ML community even apply to things that we are trying to research in HCI?''} Similarly, P13 remarked: \textit{``HCI is not about making performance better... [or] accuracy better. That is the thing that machine learning people should do. We [in HCI] have to find the problems that they have to solve.''}

Other participants expressed concern that ML/NLP reviewing norms were already shaping what HCI work gets published. P8 recounted reviewing a paper that \textit{``had literally nothing novel in the interaction''} but presented \textit{``a very extensive evaluation method''} of an LLM using standard metrics; they speculated it was \textit{``a recycled paper from an NLP venue.''} The paper was accepted, but P8 suspected the \textit{``1AC is not an expert in this field,''} and chose not to raise the issue out of fear of jeopardizing their position as a junior scholar. P10 suggested that the lowered cost of prototyping novel systems enabled by LLMs had also allowed authors with little HCI training to enter the field and review papers without nuanced knowledge of valid HCI contributions.

Senior researcher P19, however, perceived this \textit{``clash''} as a largely welcome development, provided its excesses are kept in check. They argued that HCI reviewers are not always well-equipped to assess systems with stochastic models, and as a meta-reviewer, they had deliberately invited ML scholars to review LLM-integrated system papers, but with care. They provided ML reviewers guidelines not to expect the level of rigor typical in ML/NLP venues and to focus on applied aspects instead. They also balanced reviewer expertise, ensuring HCI-trained reviewers were included and weighing disciplinary backgrounds when writing meta-reviews. 

\subsubsection{\bf Authors expressed concern that the cost of detailed technical evaluations, expected by some reviewers, could strain limited academic budgets.} \label{4-3-8}
Participants worried that reviewer expectations for technical comparisons can sometimes impose substantial financial burdens. One interviewee described being required to conduct a technical comparison for paper acceptance that cost around \$20,000 USD. %

\subsubsection{\bf Reviewers want the gist of a system and user workflow, not to be overwhelmed with configuration details.}  \label{4-3-9}
Ten participants noted that, while it is important to provide details about LLM components like prompts, authors should communicate them concisely in the main text. %
End-to-end workflow diagrams were frequently mentioned as a useful strategy, giving reviewers a user's perspective while also showing representative input/output examples of LLM components within the system. These diagrams were described as sketches that help readers form a concrete understanding of system functioning, with excessive details pushed to appendices or supplementary materials. P12 felt that reporting full prompts would leave reviewers \textit{``confused and overwhelmed,''} while P9 and P7 said they did not value detailed prompts when reviewing systems: \textit{``Some people will ask for like insane amount of details. Some people don't really care... %
I'm on the slack side, right? If I read it, and... I have good confidence that I can implement it, then I don't bother [authors] for more.''} --P9.

\subsubsection{\bf Reviewers want justification for why authors chose to use LLMs.}  \label{4-3-10}
Ten participants mentioned that reviewers often expect authors to justify their use of LLMs, or recommended this as a guideline for writing system papers (\textit{``for me personally, I really want to know the justification of why an LLM is needed,''} P1). At the same time, participants noted that such expectations could become excessive and risk distracting from a paper’s core contributions. P10 and P13 described cases where papers were rejected because reviewers doubted whether the LLM was truly necessary. P5 reflected that these requests sometimes served mainly to reassure readers who had not yet developed an implicit trust~\cite{passi_trust_data_science} in LLMs, rather than to meaningfully assess the work. They described one instance at CSCW where responding to reviewer demands meant devoting significant time and space to defending usage, which they felt distracted from their core contributions. %

\subsubsection{\bf Reviewers make paper rejection decisions based partly on classifying work as an ``LLM wrapper,'' yet struggle to define this term consistently.}  \label{4-3-11}

In discussions of peer review, nearly all participants raised the term \textit{``LLM wrapper''}, always in the context of rejection or disparagement. Importantly, we were careful to not mention this term ourselves during the interviews unless participants used it themselves. While we collected only a small number of review sets, the term appeared in two, both written by 1ACs or 2ACs---members of the program committee.\footnote{
One of us also had personally experienced one case in a program committee meeting where a paper's acceptance hinged upon a question of whether the paper was an LLM wrapper; the unspoken argument that, if it was, the committee should reject it. The debate then centered on whether the paper had a sufficient technical contribution.}
In one case, the term was invoked to argue for acceptance (\textit{``the paper goes beyond an LLM wrapper''}). In another, it was successfully used to argue for rejection, with the 1AC noting this point as a major hinge of discussion. In this context, ``wrapper'' implied the system was easy to build and thus that the technical contribution was insufficient. %

When asked to define what constitutes an LLM wrapper, however, participants offered strikingly inconsistent responses, from the expansive (\textit{``any system that integrates an LLM''} --P3), evaluative (\textit{``a system that solves a problem that is already solved''} --P13; \textit{``wrap everything previously done by humans into a system''} --P12), motivational (\textit{``an application where the use of LLMs is not sufficiently motivated... there's no need to include an LLM, but it exists''} --P4), and interactional (\textit{``the most boring system ever... it's a chat like, not even interesting in terms of interaction''} --P7; \textit{``a type of system where the interaction is not novel or trivial... the key novelty is that they applied an LLM to a specific domain''} --P10). %
Some collapsed back into tautology: \textit{``a system that is a wrapper around an LLM... so very little, if nothing around the LLM''} --P9.

Although the definitions were inconsistent, together they revealed four recurring patterns. First, reviewers often used the term when they felt a paper had not sufficiently positioned itself in prior literature or justified its use of LLMs. Second, participants noted that simply solving a problem with an LLM is no longer enough to justify a substantive contribution, and authors must do more work to claim a contribution, especially a technical one. Third, several participants invoked a sense of ``thinness'' (P9, P11, P19), where a wrapper system was perceived as too easy or trivial to build. P3 worried this stigma led reviewers to dismiss otherwise valuable insights into user behavior. Finally, participants pointed to \textit{interface resemblance} with ChatGPT: systems with chatbot-like UIs were more likely to be labeled wrappers. Two participants working on chatbots (P1, P11) felt their work was uniquely difficult to publish, as reviewers assumed little effort had gone into system design. In response, some authors reported deliberately avoiding chatbot-style interactions to avoid the ``wrapper'' stigma.

\subsubsection{\bf Reviewers hold authors accountable to work that was not yet publicly available.}  \label{4-3-12}
Several participants raised concerns that reviewers sometimes  can ask authors to justify their systems against pre-prints or published work unreleased at time of submission. Both P9 and P5 described such experiences, with P5 reporting that their paper was unfairly rejected because of it. In one collected review set, an external reviewer cited an unpublished arXiv pre-print to argue that a system on the same general topic already existed. %
The pre-print's approach, goals, and methodology had almost nothing to do with authors' work; instead, it represented an ML benchmarking paper rather than an HCI system contribution.

\subsubsection{\bf Hype and financial incentives around AI cause reviewers to cast suspicion on authors' claims of performance, generalizability, or motivation.}  \label{4-3-13}

Several participants suspected ulterior motives behind AI system papers. P4 felt that some authors included LLMs merely to be \textit{``on trend,''} while P10 described how students in particular are incentivized economically and socially to foreground LLM usage: either to signal competence for lucrative industry jobs post-graduation, or to attract visibility on social media. The latter could then result in citations and notoriety that further reinforce these incentives.  

Participants also often expressed fatigue with the hype around AI, especially on social media, and worried that some authors over-generalize claims of AI performance for attention. This concern intersected with critiques of technosolutionist rhetoric~\cite{toyama2015geek}---the notion that AI can ``save'' society. For instance, P9 emphasized wanting more \textit{``productive conversations about LLMs, before we start to posit about how we're going to save everyone else.''}

\subsubsection{\bf Non-LLM system components face dismissal under the question, ``can't an AI do that?''} \label{4-3-4}  
Related to the challenge of justifying novelty \textit{with} an LLM, participants noted that \textit{non-LLM} aspects of systems can also face skepticism. Several described situations where reviewers downplayed technical contributions or complex architectures by suggesting that an LLM could achieve the same result, either now or in the near future. This uncertainty about the continued relevance of technical work left authors struggling to demonstrate the importance of their design choices.  

Two more senior researchers expressed concern that junior reviewers (who may not remember HCI before the advent of LLMs) were particularly prone to dismissing non-LLM contributions with unverified claims such as ``an LLM could do that.'' They worried that such reviewers may also lack the training to assess technical sections or evaluations that extend beyond standard text generation benchmarks. P18, for example, reported that although their system included an LLM, reviewers challenged the \textit{non-LLM} components by asking why so much technical effort was necessary when LLMs already \textit{``do so much.''} Interestingly, this skepticism was echoed by one junior reviewer in our study, who admitted confusion about why some authors did not simply use an LLM rather than pursue what they perceived as \textit{``overly complex''} architectures.

\subsubsection{\bf Problems in peer review of LLM papers perceived as part of broader systemic issues and changes in HCI}  \label{4-3-14}
Assistant professor P10, reflecting on systems research before LLMs, emphasized that problems around reviewing LLM papers are part of broader systemic ``supply and demand'' challenges in HCI peer review, which LLMs have merely accelerated. With rising submission numbers and reviewer workload at venues like CHI and UIST, they perceived that reviewers---especially more junior scholars or those from non-HCI backgrounds---increasingly evaluate papers less on the merits of authors' claimed contributions and the appropriateness of their methods, and rather whether default ``laundry list'' of expected sections is present: specifically, a formative study, technical evaluation, and comparative usability study. This trend, they argued, ignores long-standing efforts in HCI to help reviewers appreciate diverse methodologies~\cite{greenberg2008usability}. P5 similarly critiqued the pressure to conduct formative studies, calling the expectation \textit{``ridiculous''} and elaborating: \textit{``I hate that because... my design decision is basically derived from my personal experiences.''} %

\subsubsection{\bf Guidelines needed with care and context}  \label{4-3-15}
The majority of authors felt that clearer standards are sorely needed for LLM-integrated system papers, echoing Crisan's observation that the HCI community has not yet established standards to evaluate such work~\cite{crisan2024we}. At the same time, many participants expressed hesitation about formalizing rigid guidelines. Some worried that rigid rules would unfairly constrain the freedom of authors and reviewers, given the expansive scope of possible contributions in HCI. Others argued that venue-specific communities should collectively define their own expectations, ideally with concrete examples of valid contributions. Still others felt that systemic forces---such as overburdened program committees or the influx of reviewers with limited HCI training---could not be solved through guidelines alone.  

When participants suggested guidelines, their recommendations ranged widely. Some called for more open reviewing practices (P5, P12), others for including a small-scale technical evaluation (P11--13, P15--16), and still others for reporting mistakes or potential harms and misuses (P4, P6, P8, P11--12, P14, P16). A few also emphasized the importance of training junior scholars in reviewing (P7, P10). 

Despite this lack of consensus, a set of emergent best practices can be identified by examining authors' learned strategies alongside their peer review experiences.
In Section~\ref{recommendations}, we synthesize these findings to construct initial considerations for authors, reviewers, and the HCI community, as a summary of participants' suggestions and expectations as reviewers.

\section{Considerations for Authors, Reviewers, and the HCI Community}  \label{recommendations}

Here we present a set of considerations and suggestions for authors, reviewers, and HCI communities, concerning papers that report LLM-integrated systems. 

\textbf{Methodology.} We initially synthesized this section from four sources of information in our findings: participant' learned strategies as authors; their preferences as reviewers; perceived challenges in peer review (4.3); and direct guidelines suggested across multiple participants. We also designed the tone of this section to address participant concerns over format and usability (4.3.15)---opting for a softer framing of ``considerations'' and ``suggestions,'' rather than a checklist, as well as presenting outcomes in simple list, versus more complex formats. 
Table~\ref{tab:traceability_matrix} shows how each reporting guideline connects to the subsection(s) of our findings. %

Given the limitations of our study pool, %
we sought broader validation from researchers. We reached out to thirty (30) researchers for feedback who had at least 15 years experience in HCI, who hold prominent positions in HCI community, including serving on past organizing committees of CHI and UIST, and have earned awards and recognition from ACM. Of these 30, six (6) researchers graciously volunteered their time, who hold on average about 20 years' experience each. Four of these researchers agreed to be de-anonymized, and are listed in our Acknowledgments. We shared our considerations through a Google Doc alongside a pre-print of this paper for context, and asked them to provide feedback asynchronously and/or through a virtual meeting with a note-taker present. Of these six (6), three (3) met with us synchronously. We compiled individual feedback to arrive at common themes of changes and concerns, and amended this section in light of their feedback. Reviewers of this paper also requested softening the tone of this section. Originally, we had adopted a more imperative tone, but the version presented here is more contextualized with rationales from our findings. For an alternative version of our reporting guidelines that speaks more directly to authors on ``what to do,'' see our \href{https://ianarawjo.github.io/Guidelines-for-Reporting-LLM-Integrated-Systems-in-HCI/}{\textcolor{blue}{companion website}}. HCI scholars are also invited to comment by opening a Discussion on 
\href{https://github.com/ianarawjo/Guidelines-for-Reporting-LLM-Integrated-Systems-in-HCI/discussions}{\textcolor{blue}{our public GitHub page}}. 

\subsection{Guidelines for Authors Reporting LLM-Integrated Systems in HCI} \label{sec:authors}

\textit{The centrality of LLM component(s) to authors' claimed contributions dictates the ``weight'' of reporting guidelines.} %
Papers reporting systems where LLMs are peripheral features that are easy to engineer, such as simple summary buttons, need not consider the same depth of reporting (e.g., might exclude technical evaluations). In Appendix~\ref{appendix-examples}, we provide two concrete examples of how centrality to claimed contributions affects the degree of rigor expected. %

\paragraph{\textbf{Transparency regarding prompts and configuration}} Participants in our study expected the transparent reporting of prompts and configuration details that were central to author claims, or were otherwise difficult to engineer, rather than a universal standard of reporting all prompts. The expected rigor of reporting efforts was mediated by the sensitivity of the problem domain---high-stakes domains like health or finance warranted more comprehensive reporting. Reviewers also felt that reading prompts helped them to validate systems and author effort. The appendix was considered a good place to put prompts, especially when they were lengthy. Finally, participants stressed that authors need to report the exact model name and version~(consistent with past guidelines~\cite{pangLLMificationCHI2025}). 

Based on these findings, authors should prioritize reporting prompts and technical details that directly affect reader's ability to validate core claims around system or user behavior. Non-critical components (e.g., a text summarization button) need not be so exhaustively documented, although they might remark that these components were straightforward to engineer. Exact model names should be reported whenever possible. Faced with word limits, authors might place details in appendices or supplementary material. %

\paragraph{\textbf{Visualizing system architecture and data flow}} Reviewers wanted to gain a concrete sense of how LLM components integrate into a larger system,  without being overwhelmed by low-level details. To achieve this, they pointed to practices such as including \textit{representative input-output examples} or \textit{prompt sketches}, which convey LLM component design, input structure, and output behavior. They referenced prior work that effectively presents these materials using data-flow diagrams or tables, pairing prompts with example inputs and outputs. Authors should consider how to provide a concise, concrete account of LLM component behavior and how these components integrate into the overall system.

\paragraph{\textbf{Evaluating LLM components central to claims}}
Participants felt that technical evaluations, outside of user studies, were increasingly expected for LLM components central to author claims. As reviewers, technical evaluations helped them to contextualize the robustness and generalizability of authors' systems in the absence of confounds; as authors, they faced situations where reviewers expected technical evaluations as a barrier to publication. When LLM components are integral to author claims, authors should thus seek ways to evaluate and report on these components' robustness and generalizability, outside of a user study.  
One common approach in recently published CHI and UIST papers~\cite{zhang2025chainbuddy, semantic_commit, shaikh2024rehearsal, zhou2025instructpipe} is to conduct a modest technical evaluation of the LLM component on a dataset of representative inputs (an ``eval''~\cite{shankar2024validates}). Datasets are typically custom-tailored to the use case and smaller than formal benchmarks. Metrics might be automated, expert ratings, or both. %

\paragraph{\textbf{Disclosing failure modes of LLM components}} When asked for guidelines, participants preferred that authors report potential mistakes, errors, or harms of LLM components in their systems. Reporting failure modes clarifies system reliability, provides transparency about the boundary conditions of claims, and aligns with ongoing efforts to provide transparency on potential negative impacts~\cite{sturdee2021consequences, hecht2021s}. Reviewers also felt authors could over-generalize their AI systems' performance to drum up attention and hype. In our reviewing experience, we have encountered authors making sweeping claims of positive performance that imply their system is infallable; such claims, when presented without evidence, understandably create skepticism in reviewers. Providing concrete examples of system failure modes and biases could help readers better situate how systems might realistically perform in real-world scenarios, and guard against over-trust in AI. %

\paragraph{\textbf{Justifying the choice to use LLMs}}

Reviewers appreciated when authors justified their use of LLMs relative to other possible technical approaches for the problem domain. This concern intersected with the pejorative ``LLM wrapper,'' which some felt associated with insufficient motivation for using an LLM or limited engagement with past literature or alternative technical approaches.
Note that demands for justification referred to LLM usage \textit{in general}, rather than a \textit{specific} model.
Justifications for choosing a specific model  could boil down to practical considerations such as granted credits, access, or latency, rather than strategic decisions about performance.
When designing and reporting LLM-integrated systems, authors should reflect on whether an LLM is necessary for their goals, articulate this motivation clearly, and situate their system within a broader space of alternative technical approaches.

\paragraph{\textbf{Documenting engineering processes of LLM components}}
Some participants valued brief descriptions of how authors engineered their LLM components, as these accounts conveyed the care and effort involved in system development and helped readers gauge the difficulty of re-implementing similar components. Authors might consider including a short ``engineering methodology'' subsection near the system implementation, outlining how LLM components were developed, tuned, and refined. Such descriptions can explain key design choices and make visible the iterative refinement processes necessitated by LLM uncertainty.

\paragraph{\textbf{Framing the contribution to HCI beyond the application of LLMs}}
Some authors deliberately de-emphasized references to LLMs or AI in their titles, abstracts, and introductions.
This strategy helped center HCI contributions that extend beyond the increasingly widespread application of LLMs  (Fig.~\ref{fig:chi-uist-counts}), reduce reviewer fatigue with AI-focused papers~\cite{crisan2024we}, and mitigate the risk that reviewers fixate on LLM usage at the expense of engaging with the paper's primary contributions. %
For example, authors might reconsider whether terms like ``AI-powered'' meaningfully add to titles such as ``Supporting Doctor–Patient Trust with an AI-powered System for Overcoming Hesitation in Clinical Communication.''

\paragraph{\textbf{Future-proofing contributions against advances in AI}}
In our interviews, authors described adopting framing strategies to ``future-proof'' their contributions and mitigate reviewer concerns that rapid advances in AI models could render their work irrelevant. Three strategies were especially prominent: foregrounding a novel interaction design or paradigm as the primary contribution; positioning systems as probes to study user behavior; and engaging with problem domains that have received limited attention in HCI. Senior researchers similarly emphasized considering the longevity of contributions when pursuing LLM-integrated systems research. Authors should consider whether their claimed contributions are likely to remain meaningful as AI capabilities evolve.

\subsection{Considerations for Reviewers}
\label{sec:reviewers}

The uncertainty inherent to LLMs presents unique challenges to reviewers. To overcome heightened uncertainty, reviewers enter peer review generally requiring more details from authors of LLM-integrated systems in order to trust their claims, compared to prior, more deterministic code-based systems (Table~\ref{tab:uncertainty}). Section~\ref{sec:authors} provided a set of practical guidelines for authors to increase reporting transparency and clarity around LLM component design, behavior, and claimed contributions. When skeptical of authors' systems or claims, reviewers might find the discussion and practical suggestions there insightful.

When it comes to more direct guidance for reviewers, however, our participants were split. They generally worried about ``too strict'' guidelines restricting reviewer freedom. However, they did want to raise awareness of some common pitfalls and guard against potentially unfair behavior---especially reviewers seemingly entering HCI, whether from ML/NLP or as junior scholars, and appearing to bring some mismatched norms and expectations into peer review. %
Thus, we present a few considerations for reviewers, aimed at helping reviewers evaluate HCI work fairly while respecting the diversity of HCI contribution types and epistemological traditions.

\paragraph{\textbf{Tempering expectations for rigorous technical evaluations.}}
Participants noted that expectations for extensive technical evaluation are sometimes misaligned with the goals of HCI research, where contributions frequently center on interaction paradigms, proofs-of-concept, design explorations, or user understanding rather than benchmarking model behavior. They also cautioned that requests for large-scale or comparative LLM evaluations (e.g., across multiple models or baselines) could impose substantial financial costs.
Reviewers should consider calibrating expectations for technical evaluations with authors' core claims, and avoid defaulting to demands for extensive quantitative analyses or benchmarking when such evaluations are not necessary to validate authors' stated contributions.%

\paragraph{\textbf{Contextualizing expectations for LLM justification}}
Participants observed that requests for justification for LLM use often vary in appropriateness depending on the goals and context of the paper. In some cases (e.g., privacy-sensitive or safety-critical applications), probing the rationale for using an LLM may be important. In others, however, demands to demonstrate that a problem was ``hard enough'' to warrant LLM use were viewed as misaligned with HCI goals, where rapid prototyping and exploratory system building are often central. Reviewers might consider whether and how requests for justification are proportionate to the role of the LLM in the contribution, and avoid blanket expectations regarding model choice, including assumptions that only open models are acceptable~\cite{baltes2025evaluation}, when proprietary models reasonably support research aims.

\paragraph{\textbf{``LLM wrapper'' may obscure rather than clarify.}} Our participants used the term ``LLM wrapper'' when considering whether to accept or reject research for publication. In two review sets we collected, the term ``LLM wrapper'' also appeared in key arguments made by reviewers. Yet, our participants held very inconsistent, subjective understandings of this term. This variability suggests that use of LLM wrapper or similar trendy terms in peer review may obscure rather than clarify reasoning, creating an illusion of shared understanding while diverting attention from substantive concerns. The term also risks conflating judgments about the difficulty of building a system with assessments of overall contribution quality. Reviewers should consider articulating their specific concerns directly, rather than relying on this label.

\paragraph{\textbf{Respecting non-LLM systems and minimizing reductive comparisons.}}
Participants cautioned that dismissive comparisons (e.g., suggesting that ``an LLM could do that'') risk trivializing technical and design contributions that pursue different research questions, design goals, or ethical commitments. Authors may deliberately avoid LLMs for a range of legitimate reasons, including methodological, ethical, or political considerations. Reviewers might take care to evaluate non-LLM systems on their own terms, and to afford them the same respect as work that integrates LLMs.

\paragraph{\textbf{Applying a contextual notion of reproducibility}} Venues for more technical computing fields such as NLP and SE are increasingly requiring comparison to open models and code submissions alongside LLM-integrated work~\cite{baltes2025evaluation}. By contrast, our participants emphasized that reproducibility expectations are not uniform across HCI. The value of HCI research often lies in inspiring new interaction paradigms or understanding users, rather than enabling exact replication under tightly controlled conditions. 
For example, Park et al.'s work on generative agents~\cite{park2023generative} led to substantial excitement and follow-up work---inspiration that holds regardless of whether its core simulation can be precisely replicated. For LLM-integrated systems, reviewers should consider continuing this ``soft'' notion of reproducibility consistent with longstanding HCI norms~\cite{brown2011intothewild}.

\paragraph{\textbf{Familiarizing oneself with the diversity of HCI contribution types}}
Participants suggested that reviewer inconsistency partly stems from differing expectations about what constitutes a valid contribution and level of methodological rigor. Reviewers with limited experience in HCI would benefit from familiarizing themselves with the range of contribution types accepted at HCI venues, and how these differ from the methodological and value commitments of more technically oriented  fields. Prior work by Wobbrock and Kientz~\cite{wobbrock2016research} outlines common HCI contribution types, while Soden et al.~\cite{soden2024evaluating} and Crabtree~\cite{crabtree2025h} describe qualitative and interpretive methodologies frequently used in HCI. %

\paragraph{\textbf{Ensure balanced HCI expertise on review panels}}
Authors judged that external reviewers from more technical domains could overly focus on technical contributions to the detriment of overlooking other HCI contributions; these reviewers' perspectives could then dominate peer review. A modest suggestion here is for meta-reviewers and area chairs to try to ensure at least one external reviewer has expertise in HCI. Balance is essential to ensure that non-technical contributions are fairly assessed and weighted; reviewers from more technical domains can provide valuable insights, but their perspectives should complement rather than dominate the evaluation of HCI work.

\subsection{Suggestions for the HCI community}

Finally, we offer broader suggestions for the HCI community, based on  suggestions made across multiple participants. Note that we do not include making better use of desk reject~\cite{crisan2024we}, since CHI and UIST have already adopted this practice.

\paragraph{\textbf{Facilitate venue-specific discussions to establish guidelines for papers that report LLM-integrated systems.}} Beyond the baseline fluctuations of peer review, authors felt reviewers applied inconsistent standards to this paper type, whether due to differences in the norms of their disciplinary backgrounds (e.g., HCI vs. ML/NLP), personal and moral beliefs, or past experience using LLMs. 
To better set shared norms, HCI venues or committees (e.g., CHI, UIST, CSCW) might convene focused discussions or workshops among experienced members to define how LLM-integrated systems should be reported  within their context. These conversations could result in publicly available guidelines (posted on conference websites, blogs, or linked from Calls for Papers) so that expectations are transparent to both authors and reviewers.%

\paragraph{\textbf{Curate paper exemplars that follow good practices for reporting.}}
In the absence of concrete guidelines, authors looked to recently published papers for inspiration in how to report LLM-integrated systems. %
Venues might curate a set of exemplary papers demonstrating good practices, along with anonymized or composite examples of common pitfalls. Such exemplars can serve as teaching materials for authors and reviewers, illustrating what each venue considers a strong contribution. Examples of LLM-integrated system papers in HCI that reflect many of our reporting considerations are: Rehearsal~\cite{shaikh2024rehearsal}, BloomIntent~\cite{choi2025bloomintent}, EvAlignUX~\cite{zheng2025evalignux}, GUM~\cite{shaikh2025gum},  ChainBuddy~\cite{zhang2025chainbuddy}, and InstructPipe~\cite{zhou2025instructpipe}. 

\paragraph{\textbf{Consider adopting open reviewing practices.}} When asked for guidelines, multiple participants felt that posting anonymized reviews after  submission cycles would better contribute to open science, help authors better understand reviewer expectations, help communities track evolving standards, and allow outsiders to decide themselves on the presence or absence of potentially AI-generated reviews.\footnote{Since writing this recommendation, there was controversy over AI-generated reviews at ICLR 2026. The community mobilized to map the scope of the problem (e.g., \url{https://iclr.pangram.com/reviews})---possible because ICLR makes anonymized reviews public.}

\paragraph{\textbf{Consider requiring submission of code or equivalent supplementary materials.}}
Participants described growing tension between the need for additional transparency when reporting LLM-integrated systems, and page and word limits. Reviewers also felt that inspecting prompts directly helped them assess system quality and author effort. Venues could mitigate this tension by going beyond encouragement to requiring submission of code alongside system papers. Authors could mark LLM-related resources clearly (e.g., a dedicated \texttt{prompts/} folder) to facilitate review. For example, HCI venues might look to the ACM artifact badging framework~\cite{ACM2020} as implemented in fields like software engineering. When code release is infeasible, supplementary material can serve the same purpose. Public release is ideal, but providing code privately for review could be sufficient to improve transparency and trust.

\section{Discussion} %

Three major themes emerged from our findings. First, \textit{the uncertainty of LLMs appears to erode trust between authors, their systems, and reviewers}, exceeding prior norms of trust-building in HCI research and peer review. Authors have learned strategies to re-establish trust in the face of this uncertainty, whether with reviewers, or with faith in their own system. %
Reviewers, in turn, have reacted to the uncertainty and hype around LLMs by demanding more extensive details and evaluations, judging that a user study alone with positive user feedback no longer suffices to establish credibility for publication. %
Authors also perceived that the acceptance bar for LLM-integrated system papers has risen. %
When considered alongside perceptions that LLMs have lowered the barrier to prototyping, what was once the \textit{burden of execution}---taking time and effort to implement a system---seems to be shifting to the \textit{burden of validation}---expecting verifications and triangulations of system performance~\cite{subramonyam2023bridging}. Yet, even as reviewers request more validation and detail, they still expect adherence to pre-LLM constraints on page length and word count, creating tension that authors often resolve by moving content into appendices or code repositories. %

Second, interviewees noted that the ongoing hype around LLMs has led both authors and reviewers to treat them as uniquely special, when they are increasingly just implementation details. For authors, this hype can incentivize foregrounding LLMs in titles, abstracts, or paper framing to signal novelty, timeliness, fundability, or improve employment prospects. For reviewers, it can result in disproportionate focus on the LLM component: overlooking other contributions, demanding excessive technical evidence, questioning the rationale for using LLMs, or engaging in moral gatekeeping over whether such use is appropriate. %
Reviewers' judgments often hinged on whether a system was seen as an ``LLM wrapper,'' a label whose meaning varied widely across participants. %

Third, many tensions and perceptions of reviewer inconsistency appeared to stem from \textit{a clash of communities and their values and expectations}, particularly between HCI and more technical fields such as NLP and ML. 
Authors reported that reviewers with limited HCI experience increasingly apply the evaluative norms of their home fields, emphasizing technical rigor and performance metrics over interaction design or user insights. Notably, the more senior researchers we spoke to, including one volunteering editor of our considerations section, also echoed this concern.
This clash has produced two shifts: an expectation that any LLM-integrated system must include a technical evaluation beyond a user study, and a perception that some reviewers undervalue core HCI contributions. %
Authors questioned whether calls for additional technical rigor always serve constructive ends. %
Practices providing review guidelines to support ML reviewers and balancing reviewer pools to ensure HCI perspectives are represented (as suggested by P19) offer concrete ways to mitigate these tensions while preserving the benefits of interdisciplinary exchange.

Lastly, while we did our best to prepare a set of considerations and strengthen them with external feedback, we recognize that our work is but one part of a larger, ongoing conversation. In the future, HCI communities may seek to collectively establish more solid guidelines for LLM-integrated systems, particularly as systems grow more complex. We hope our work serves as a productive starting point for future discussions and debates around this topic. %

\subsection{Connections and Future Work}

\subsubsection{Epistemological clashes in HCI peer review}

As discussed in Section~2, longstanding debates about peer review quality in HCI often stem from clashes between positivist assumptions about what constitutes scientific validity (i.e., emphasizing quantification, lab experiments, and generalization~\cite{crabtree2025h}) versus more qualitative or design-oriented approaches. %
If LLM-integrated systems attract more reviewers from ML or NLP backgrounds---a plausible scenario---these positivist expectations may continue to shape HCI reviewing norms. Junior reviewers trained in technical fields may also inherit these epistemological assumptions, as Crabtree notes that positivism remains dominant in the broader scientific culture~\cite{crabtree2025h}.

On the one hand, we want to encourage more people to join the HCI community, who bring their diverse backgrounds, perspectives, and skills, to help us together consider computing problems in a more human-centered way. HCI scholars also benefit from being challenged and expanding the reach of the field. However, growth must be sustainable, and we must ensure unique values of HCI research remain intact. It could be that the sudden interest in HCI has brought with it a greater need for education on HCI's unique values, methodologies, and history, especially how HCI approaches can differ from more traditional conceptions of science.  

\subsubsection{Comparison to other reporting guidelines and suggestions} %

Our work contributes to an emerging discourse on reporting practices for LLM-integrated systems. Some of our suggestions align with past guidelines, such as ``make a reasonable attempt to justify the use of an LLM'' or to report exact model names~\cite{pangLLMificationCHI2025, crisan2024we}. In other cases they diverge. For instance, Crisan~\cite{crisan2024we} suggests that authors should evaluate system sensitivity to prompt choice; no participant mentioned this (instead, we would recommend testing against a representative dataset). Pang et al.~\cite{pangLLMificationCHI2025}~recommend that authors use open models and disclose all prompts, %
while a pre-print in SE asks authors to always ``include an open LLM as a baseline'' and %
``publish all prompts'' that were considered across the entire development cycle~\cite{baltes2025evaluation}. Our findings show that authors in HCI have reasonable justifications for going against these guidelines, while the clash of communities suggests HCI needs its own standards. %

Our work also relates to previous attempts to provide reporting  frameworks to enhance transparency. While aware of reporting frameworks for ML artifacts like model cards and datasheets for datasets~\cite{mitchell2019model, gebru2021datasheets}, we felt that LLM-integrated systems are more varied artifacts and far from settled technology; thus, providing an ``LLM-integrated system card'' or checklist was premature. Such a card is also perhaps too prescriptive for the vast range of HCI contexts where LLM-integrated systems may appear, and could unwittingly push authors into centering LLMs/AI unnecessarily.

\subsubsection{``LLMs erode trust'' between parties}

One of the chief values of qualitative research is producing analytic insights that may be transferrable or shed light on other contexts~\cite[p. 16-7]{crabtree2025h}. Beyond peer review, our findings contribute to an analytic lens of \textit{LLMs eroding trust by creating uncertainty} in collaborative and negotiation settings. 
This lens calls researchers to attend to how the presence of LLMs may make trust-building harder among people trying to coordinate. For example, in peer review, authors now question whether reviewers used LLMs to review their paper, while reviewers question whether authors used LLMs to write significant parts of their papers. We can see this too in professor-student relationships, where the possibility that AI may have been used can put otherwise stellar student’s credibility in question where it may not have before. 

The idea that LLMs erode trust has precedence in the literature. In Passi \& Jackson's ``Trust in Data Science,'' the introduction of a ``neural network'' to model data at a business meeting caused distrust between the presenter and his audience~%
\cite[p. 16]{passi_trust_data_science}. Such distrust appeared uniquely attributable to the ``inscrutable'' nature of neural networks---the uncertainty around their outputs, operation, and capabilities---and did not occur for more ``scrutable'' models. %
To re-establish trust, data scientists argued for ``implicit trust with explicit verification''---%
having a baseline trust in neural network's power (which provides the presenter benefit of the doubt), and conducting an  ``explicit verification'' akin to a technical evaluation across representative inputs~\cite{passi_trust_data_science}. %

\subsubsection{Author's justifications for generally favoring proprietary models, versus ethical stances against usage}

We found that prior work generally treats open and small models as a public good and pits them in contrast to large, proprietary APIs~\cite{chaaben2025mlmodelselection, pangLLMificationCHI2025, kapania2025ethicsofllmuse}. From a sustainability perspective, chaaben et al. argue that ``this trend of using complex models [LLMs like ChatGPT] leads to worrying forecasts of energy consumption of data centers, AI, and cryptocurrencies''~\cite{chaaben2025mlmodelselection}. Other arguments for open models %
is that they enable reproducibility~\cite{pangLLMificationCHI2025} and avoid sudden deprecation or silent updates~\cite{ma2024schrodinger}.

While LLM's energy usage is concerning, participants had a range of justifications for favoring proprietary models in the present moment. Like Kapania et al.~\cite{kapania2025ethicsofllmuse}, we also found that HCI authors were generally knowledgeable about ethical concerns, even though they themselves had rationalizations or workarounds for not acting upon best practices. As authors ourselves, we sympathize with these justifications. %
We have used OpenAI models, for instance, in past papers in no small part due to research credits granted from OpenAI. Yet in such a climate, how can the HCI community properly incentivize and value open model usage, without making it a requirement? %
Venues that wish to reward open model usage might consider incentives such as special artifact badges or sustainability awards for papers that strictly uses open models.

\subsubsection{Need for methodologies to cite for LLM component engineering}

Authors described their rapid, iterative LLM component engineering processes, but had no way to discuss them as an ``expected'' or ``valid'' methodology. Instead, authors hesitated in interviews, seemingly worried that we would cast judgment on the apparent lack of ``scientific rigor'' of their methodology. However, we can report here that this messy iterative process is just how LLM-integrated system work is done.
Other research backs up the reasons why LLM component building is \textit{necessarily} iterative~%
\cite{shankar2024validates}. The iterative process authors report also resembles Yang et al.'s technology-driven innovation~\cite{yang2020re} and the three stages of prompt engineering Arawjo et al. introduce in prior work~\cite{arawjo2024chainforge}, although with more social validation towards the latter stages. %
To allay reviewer concerns and communicate this engineering methodology as an acceptable standard, some authors wanted  literature to cite. %
Future work might explicitly compare different LLM component engineering methodologies. %

\subsubsection{Tools to support efficient, transparent engineering and reporting}

Prior work in prompt engineering and LLM evaluation focuses on tools and techniques to help users improve prompts, choose models, and audit bias \cite{kim2024evallm, shankar2024validates, arawjo2024chainforge, gebreegziabher2025metricmate, khattab2024dspy}. However, engineering and evaluation ultimately exist within broader social and organizational contexts where evaluators need to report their findings in an accessible manner to stakeholders. A recurring theme in our conversations was not participants thinking evaluations were useless, but that they weren't worth the effort, taking time away from other, more critical research tasks. %
Better tooling, to generate reports of what authors tried and even produce figures and statistics, could improve the situation around %
transparency 
without overly burdening authors. For instance, a system could guide authors through the process of conducting a technical evaluation of an LLM component, from strategy and evaluation criteria formation \cite{shankar2024validates, gebreegziabher2025metricmate}, to comparison across model and prompt alternatives~\cite{arawjo2024chainforge, kim2024evallm}, to generating the report or report structure. %

\begin{acks}
We would like to thank Daniel Vogel (University of Waterloo), Fanny Chevalier (University of Toronto), Steve Oney (University of Michigan), and Björn Hartmann (University of California, Berkeley) for their feedback on earlier versions of our considerations section. 
\end{acks}

\bibliographystyle{ACM-Reference-Format}
\bibliography{citations}


\begin{thebibliography}{50}

%%% the Web address.

\ifx \showCODEN    \undefined \def \showCODEN     #1{\unskip}     \fi
\ifx \showISBNx    \undefined \def \showISBNx     #1{\unskip}     \fi
\ifx \showISBNxiii \undefined \def \showISBNxiii  #1{\unskip}     \fi
\ifx \showISSN     \undefined \def \showISSN      #1{\unskip}     \fi
\ifx \showLCCN     \undefined \def \showLCCN      #1{\unskip}     \fi
\ifx \shownote     \undefined \def \shownote      #1{#1}          \fi
\ifx \showarticletitle \undefined \def \showarticletitle #1{#1}   \fi
\ifx \showURL      \undefined \def \showURL       {\relax}        \fi
% invisible to TeX
\providecommand\bibfield[2]{#2}
\providecommand\bibinfo[2]{#2}
\providecommand\natexlab[1]{#1}
\providecommand\showeprint[2][]{arXiv:#2}

\bibitem[{ACM}(2020)]%
        {ACM2020}
\bibfield{author}{\bibinfo{person}{{ACM}}.} \bibinfo{year}{2020}\natexlab{}.
\newblock \bibinfo{title}{Artifact Review and Badging Version 1.1}.
\newblock \bibinfo{howpublished}{\url{https://www.acm.org/publications/policies/artifact-review-and-badging-current}}.
\newblock


\bibitem[{ACM}(2025)]%
        {acm-peer-review-faq}
\bibfield{author}{\bibinfo{person}{{ACM}}.} \bibinfo{year}{2025}\natexlab{}.
\newblock \bibinfo{title}{ACM Peer Review Policy Frequently Asked Questions}.
\newblock \bibinfo{howpublished}{\url{https://www.acm.org/publications/policies/peer-review-faq}}.
\newblock
\newblock
\shownote{Accessed: 2025-08-06}.


\bibitem[{ACM CSCW}(2025)]%
        {cscw2026_reviewprocess}
\bibfield{author}{\bibinfo{person}{{ACM CSCW}}.} \bibinfo{year}{2025}\natexlab{}.
\newblock \bibinfo{title}{Paper Review Process at {CSCW} 2026}.
\newblock \bibinfo{howpublished}{\url{https://cscw.acm.org/2026/blog/reviewprocess.html}}.
\newblock
\newblock
\shownote{Blog post, CSCW 2026 Technical Program Blog}.


\bibitem[Arawjo(2024)]%
        {arawjollmwrapper}
\bibfield{author}{\bibinfo{person}{Ian Arawjo}.} \bibinfo{year}{2024}\natexlab{}.
\newblock \bibinfo{title}{{LLM} Wrapper Papers Are Hurting {HCI} Research}.
\newblock \bibinfo{howpublished}{Medium}.
\newblock
\newblock
\shownote{Blog post}.


\bibitem[Arawjo et~al\mbox{.}(2024)]%
        {arawjo2024chainforge}
\bibfield{author}{\bibinfo{person}{Ian Arawjo}, \bibinfo{person}{Chelse Swoopes}, \bibinfo{person}{Priyan Vaithilingam}, \bibinfo{person}{Martin Wattenberg}, {and} \bibinfo{person}{Elena~L. Glassman}.} \bibinfo{year}{2024}\natexlab{}.
\newblock \showarticletitle{ChainForge: A Visual Toolkit for Prompt Engineering and {LLM} Hypothesis Testing}. In \bibinfo{booktitle}{\emph{Proceedings of the 2024 {CHI} Conference on Human Factors in Computing Systems}} \emph{(\bibinfo{series}{CHI '24})}. \bibinfo{publisher}{Association for Computing Machinery}, \bibinfo{address}{New York, NY, USA}, Article \bibinfo{articleno}{304}, \bibinfo{numpages}{18}~pages.
\newblock
\href{https://doi.org/10.1145/3613904.3642016}{doi:\nolinkurl{10.1145/3613904.3642016}}


\bibitem[Baltes et~al\mbox{.}(2025)]%
        {baltes2025evaluation}
\bibfield{author}{\bibinfo{person}{Sebastian Baltes}, \bibinfo{person}{Florian Angermeir}, \bibinfo{person}{Chetan Arora}, \bibinfo{person}{Marvin Mu{\~n}oz~Bar{\'o}n}, \bibinfo{person}{Chunyang Chen}, \bibinfo{person}{Lukas B{\"o}hme}, \bibinfo{person}{Fabio Calefato}, \bibinfo{person}{Neil Ernst}, \bibinfo{person}{Davide Falessi}, \bibinfo{person}{Brian Fitzgerald}, {et~al\mbox{.}}} \bibinfo{year}{2025}\natexlab{}.
\newblock \showarticletitle{Evaluation Guidelines for Empirical Studies in Software Engineering Involving {LLMs}}.
\newblock \bibinfo{journal}{\emph{arXiv preprint arXiv:2508.15503}} (\bibinfo{year}{2025}).
\newblock
\href{https://doi.org/10.48550/arXiv.2508.15503}{doi:\nolinkurl{10.48550/arXiv.2508.15503}}


\bibitem[Ben~Chaaben et~al\mbox{.}(2025)]%
        {chaaben2025mlmodelselection}
\bibfield{author}{\bibinfo{person}{Eya Ben~Chaaben}, \bibinfo{person}{Janin Koch}, {and} \bibinfo{person}{Wendy~E. Mackay}.} \bibinfo{year}{2025}\natexlab{}.
\newblock \showarticletitle{``Should I Choose a Smaller Model?'': Understanding {ML} Model Selection and Its Impact on Sustainability}. In \bibinfo{booktitle}{\emph{Proceedings of the 2025 {CHI} Conference on Human Factors in Computing Systems}} \emph{(\bibinfo{series}{CHI '25})}. \bibinfo{publisher}{Association for Computing Machinery}, \bibinfo{address}{New York, NY, USA}, Article \bibinfo{articleno}{1008}, \bibinfo{numpages}{13}~pages.
\newblock
\href{https://doi.org/10.1145/3706598.3713240}{doi:\nolinkurl{10.1145/3706598.3713240}}


\bibitem[Ben~Chaaben et~al\mbox{.}(2024)]%
        {syriani2024toward}
\bibfield{author}{\bibinfo{person}{Meriem Ben~Chaaben}, \bibinfo{person}{Oussama Ben~Sghaier}, \bibinfo{person}{Mouna Dhaouadi}, \bibinfo{person}{Nafisa Elrasheed}, \bibinfo{person}{Ikram Darif}, \bibinfo{person}{Imen Jaoua}, \bibinfo{person}{Bentley Oakes}, \bibinfo{person}{Eugene Syriani}, {and} \bibinfo{person}{Mohammad Hamdaqa}.} \bibinfo{year}{2024}\natexlab{}.
\newblock \showarticletitle{Toward Intelligent Generation of Tailored Graphical Concrete Syntax}. In \bibinfo{booktitle}{\emph{Proceedings of the ACM/IEEE 27th International Conference on Model Driven Engineering Languages and Systems}}. \bibinfo{pages}{160--171}.
\newblock


\bibitem[Brown et~al\mbox{.}(2011)]%
        {brown2011intothewild}
\bibfield{author}{\bibinfo{person}{Barry Brown}, \bibinfo{person}{Stuart Reeves}, {and} \bibinfo{person}{Scott Sherwood}.} \bibinfo{year}{2011}\natexlab{}.
\newblock \showarticletitle{Into the Wild: Challenges and Opportunities for Field Trial Methods}. In \bibinfo{booktitle}{\emph{Proceedings of the SIGCHI Conference on Human Factors in Computing Systems}} \emph{(\bibinfo{series}{CHI '11})}. \bibinfo{publisher}{Association for Computing Machinery}, \bibinfo{address}{New York, NY, USA}, \bibinfo{pages}{1657--1666}.
\newblock
\href{https://doi.org/10.1145/1978942.1979185}{doi:\nolinkurl{10.1145/1978942.1979185}}


\bibitem[Charmaz(2006)]%
        {charmaz2006constructing}
\bibfield{author}{\bibinfo{person}{Kathy Charmaz}.} \bibinfo{year}{2006}\natexlab{}.
\newblock \bibinfo{booktitle}{\emph{Constructing Grounded Theory: A Practical Guide Through Qualitative Analysis}}.
\newblock \bibinfo{publisher}{Sage Publications}.
\newblock


\bibitem[Choi et~al\mbox{.}(2025)]%
        {choi2025bloomintent}
\bibfield{author}{\bibinfo{person}{Yoonseo Choi}, \bibinfo{person}{Eunhye Kim}, \bibinfo{person}{Hyunwoo Kim}, \bibinfo{person}{Donghyun Park}, \bibinfo{person}{Honggu Lee}, \bibinfo{person}{Jin~Young Kim}, {and} \bibinfo{person}{Juho Kim}.} \bibinfo{year}{2025}\natexlab{}.
\newblock \showarticletitle{BloomIntent: Automating Search Evaluation with {LLM}-Generated Fine-Grained User Intents}. In \bibinfo{booktitle}{\emph{Proceedings of the 38th Annual {ACM} Symposium on User Interface Software and Technology}} \emph{(\bibinfo{series}{UIST '25})}. \bibinfo{publisher}{Association for Computing Machinery}, \bibinfo{address}{New York, NY, USA}, Article \bibinfo{articleno}{177}, \bibinfo{numpages}{34}~pages.
\newblock
\href{https://doi.org/10.1145/3746059.3747677}{doi:\nolinkurl{10.1145/3746059.3747677}}


\bibitem[Crabtree(2025)]%
        {crabtree2025h}
\bibfield{author}{\bibinfo{person}{Andy Crabtree}.} \bibinfo{year}{2025}\natexlab{}.
\newblock \showarticletitle{H is for Human and How (Not) to Evaluate Qualitative Research in {HCI}}.
\newblock \bibinfo{journal}{\emph{Human--Computer Interaction}} (\bibinfo{year}{2025}).
\newblock
\href{https://doi.org/10.1080/07370024.2025.2475743}{doi:\nolinkurl{10.1080/07370024.2025.2475743}}
\newblock
\shownote{Advance online publication}.


\bibitem[Crisan(2024)]%
        {crisan2024we}
\bibfield{author}{\bibinfo{person}{Anamaria Crisan}.} \bibinfo{year}{2024}\natexlab{}.
\newblock \showarticletitle{We Don’t Know How to Assess {LLM} Contributions in {VIS}/{HCI}}. In \bibinfo{booktitle}{\emph{2024 {IEEE} Evaluation and Beyond---Methodological Approaches for Visualization ({BELIV})}}. \bibinfo{pages}{115--118}.
\newblock
\href{https://doi.org/10.1109/BELIV64461.2024.00018}{doi:\nolinkurl{10.1109/BELIV64461.2024.00018}}


\bibitem[Gebreegziabher et~al\mbox{.}(2025)]%
        {gebreegziabher2025metricmate}
\bibfield{author}{\bibinfo{person}{Simret~Araya Gebreegziabher}, \bibinfo{person}{Charles Chiang}, \bibinfo{person}{Zichu Wang}, \bibinfo{person}{Zahra Ashktorab}, \bibinfo{person}{Michelle Brachman}, \bibinfo{person}{Werner Geyer}, \bibinfo{person}{Toby Jia-Jun Li}, {and} \bibinfo{person}{Diego G{\'o}mez-Zar{\'a}}.} \bibinfo{year}{2025}\natexlab{}.
\newblock \showarticletitle{MetricMate: An Interactive Tool for Generating Evaluation Criteria for {LLM}-as-a-Judge Workflow}. In \bibinfo{booktitle}{\emph{Proceedings of the 4th Annual Symposium on Human-Computer Interaction for Work}}. \bibinfo{pages}{1--18}.
\newblock
\href{https://doi.org/10.1145/3729176.3729199}{doi:\nolinkurl{10.1145/3729176.3729199}}


\bibitem[Gebru et~al\mbox{.}(2021)]%
        {gebru2021datasheets}
\bibfield{author}{\bibinfo{person}{Timnit Gebru}, \bibinfo{person}{Jamie Morgenstern}, \bibinfo{person}{Briana Vecchione}, \bibinfo{person}{Jennifer Wortman~Vaughan}, \bibinfo{person}{Hanna Wallach}, \bibinfo{person}{Hal Daum{\'e}~III}, {and} \bibinfo{person}{Kate Crawford}.} \bibinfo{year}{2021}\natexlab{}.
\newblock \showarticletitle{Datasheets for Datasets}.
\newblock \bibinfo{journal}{\emph{Commun. ACM}} \bibinfo{volume}{64}, \bibinfo{number}{12} (\bibinfo{year}{2021}), \bibinfo{pages}{86--92}.
\newblock
\href{https://doi.org/10.1145/3458723}{doi:\nolinkurl{10.1145/3458723}}


\bibitem[Greenberg and Buxton(2008)]%
        {greenberg2008usability}
\bibfield{author}{\bibinfo{person}{Saul Greenberg} {and} \bibinfo{person}{Bill Buxton}.} \bibinfo{year}{2008}\natexlab{}.
\newblock \showarticletitle{Usability Evaluation Considered Harmful (Some of the Time)}. In \bibinfo{booktitle}{\emph{Proceedings of the SIGCHI Conference on Human Factors in Computing Systems}} \emph{(\bibinfo{series}{CHI '08})}. \bibinfo{publisher}{Association for Computing Machinery}, \bibinfo{address}{New York, NY, USA}, \bibinfo{pages}{111--120}.
\newblock
\href{https://doi.org/10.1145/1357054.1357074}{doi:\nolinkurl{10.1145/1357054.1357074}}


\bibitem[Hecht et~al\mbox{.}(2021)]%
        {hecht2021s}
\bibfield{author}{\bibinfo{person}{Brent Hecht}, \bibinfo{person}{Lauren Wilcox}, \bibinfo{person}{Jeffrey~P. Bigham}, \bibinfo{person}{Johannes Sch{\"o}ning}, \bibinfo{person}{Ehsan Hoque}, \bibinfo{person}{Jason Ernst}, \bibinfo{person}{Yonatan Bisk}, \bibinfo{person}{Luigi De~Russis}, \bibinfo{person}{Lana Yarosh}, \bibinfo{person}{Bushra Anjum}, {et~al\mbox{.}}} \bibinfo{year}{2021}\natexlab{}.
\newblock \showarticletitle{It’s Time to Do Something: Mitigating the Negative Impacts of Computing Through a Change to the Peer Review Process}.
\newblock \bibinfo{journal}{\emph{arXiv preprint arXiv:2112.09544}} (\bibinfo{year}{2021}).
\newblock
\href{https://doi.org/10.48550/arXiv.2112.09544}{doi:\nolinkurl{10.48550/arXiv.2112.09544}}


\bibitem[Iarygina et~al\mbox{.}(2025)]%
        {IARYGINA2025103379}
\bibfield{author}{\bibinfo{person}{Olga Iarygina}, \bibinfo{person}{Kasper Hornb{\ae}k}, {and} \bibinfo{person}{Aske Mottelson}.} \bibinfo{year}{2025}\natexlab{}.
\newblock \showarticletitle{Demand Characteristics in Human--Computer Experiments}.
\newblock \bibinfo{journal}{\emph{International Journal of Human-Computer Studies}}  \bibinfo{volume}{193} (\bibinfo{year}{2025}), \bibinfo{pages}{103379}.
\newblock
\href{https://doi.org/10.1016/j.ijhcs.2024.103379}{doi:\nolinkurl{10.1016/j.ijhcs.2024.103379}}


\bibitem[Kapania et~al\mbox{.}(2025)]%
        {kapania2025ethicsofllmuse}
\bibfield{author}{\bibinfo{person}{Shivani Kapania}, \bibinfo{person}{Ruiyi Wang}, \bibinfo{person}{Toby Jia-Jun Li}, \bibinfo{person}{Tianshi Li}, {and} \bibinfo{person}{Hong Shen}.} \bibinfo{year}{2025}\natexlab{}.
\newblock \showarticletitle{``I’m Categorizing {LLM} as a Productivity Tool'': Examining Ethics of {LLM} Use in {HCI} Research Practices}.
\newblock \bibinfo{journal}{\emph{Proceedings of the {ACM} on Human-Computer Interaction}} \bibinfo{volume}{9}, \bibinfo{number}{CSCW2}, Article \bibinfo{articleno}{CSCW102} (\bibinfo{year}{2025}), \bibinfo{numpages}{26}~pages.
\newblock
\href{https://doi.org/10.1145/3711000}{doi:\nolinkurl{10.1145/3711000}}


\bibitem[Khattab et~al\mbox{.}(2024)]%
        {khattab2024dspy}
\bibfield{author}{\bibinfo{person}{Omar Khattab}, \bibinfo{person}{Arnav Singhvi}, \bibinfo{person}{Paridhi Maheshwari}, \bibinfo{person}{Zhiyuan Zhang}, \bibinfo{person}{Keshav Santhanam}, \bibinfo{person}{Saiful Haq}, \bibinfo{person}{Ashutosh Sharma}, \bibinfo{person}{Thomas~T. Joshi}, \bibinfo{person}{Hanna Moazam}, \bibinfo{person}{Heather Miller}, {et~al\mbox{.}}} \bibinfo{year}{2024}\natexlab{}.
\newblock \showarticletitle{{DSPy}: Compiling Declarative Language Model Calls into State-of-the-Art Pipelines}. In \bibinfo{booktitle}{\emph{The Twelfth International Conference on Learning Representations}}.
\newblock
\newblock
\shownote{OpenReview preprint}.


\bibitem[Kim et~al\mbox{.}(2024)]%
        {kim2024evallm}
\bibfield{author}{\bibinfo{person}{Tae~Soo Kim}, \bibinfo{person}{Yoonjoo Lee}, \bibinfo{person}{Jamin Shin}, \bibinfo{person}{Young-Ho Kim}, {and} \bibinfo{person}{Juho Kim}.} \bibinfo{year}{2024}\natexlab{}.
\newblock \showarticletitle{{Evallm}: Interactive Evaluation of Large Language Model Prompts on User-Defined Criteria}. In \bibinfo{booktitle}{\emph{Proceedings of the 2024 {CHI} Conference on Human Factors in Computing Systems}}. \bibinfo{publisher}{Association for Computing Machinery}, \bibinfo{address}{New York, NY, USA}, \bibinfo{pages}{1--21}.
\newblock
\href{https://doi.org/10.1145/3613904.3642216}{doi:\nolinkurl{10.1145/3613904.3642216}}


\bibitem[Kloft et~al\mbox{.}(2024)]%
        {ai_nocebo_effects}
\bibfield{author}{\bibinfo{person}{Agnes~Mercedes Kloft}, \bibinfo{person}{Robin Welsch}, \bibinfo{person}{Thomas Kosch}, {and} \bibinfo{person}{Steeven Villa}.} \bibinfo{year}{2024}\natexlab{}.
\newblock \showarticletitle{``{AI} Enhances Our Performance, I Have No Doubt This One Will Do the Same'': The Placebo Effect Is Robust to Negative Descriptions of {AI}}. In \bibinfo{booktitle}{\emph{Proceedings of the 2024 {CHI} Conference on Human Factors in Computing Systems}} \emph{(\bibinfo{series}{CHI '24})}. \bibinfo{publisher}{Association for Computing Machinery}, \bibinfo{address}{New York, NY, USA}, Article \bibinfo{articleno}{299}, \bibinfo{numpages}{24}~pages.
\newblock
\href{https://doi.org/10.1145/3613904.3642633}{doi:\nolinkurl{10.1145/3613904.3642633}}


\bibitem[Kosch et~al\mbox{.}(2023)]%
        {ai_placebo_effects}
\bibfield{author}{\bibinfo{person}{Thomas Kosch}, \bibinfo{person}{Robin Welsch}, \bibinfo{person}{Lewis Chuang}, {and} \bibinfo{person}{Albrecht Schmidt}.} \bibinfo{year}{2023}\natexlab{}.
\newblock \showarticletitle{The Placebo Effect of Artificial Intelligence in Human--Computer Interaction}.
\newblock \bibinfo{journal}{\emph{{ACM} Transactions on Computer-Human Interaction}} \bibinfo{volume}{29}, \bibinfo{number}{6}, Article \bibinfo{articleno}{56} (\bibinfo{year}{2023}), \bibinfo{numpages}{32}~pages.
\newblock
\href{https://doi.org/10.1145/3529225}{doi:\nolinkurl{10.1145/3529225}}


\bibitem[Landay(2009)]%
        {landay2009_giveup_on_chiuist}
\bibfield{author}{\bibinfo{person}{James Landay}.} \bibinfo{year}{2009}\natexlab{}.
\newblock \bibinfo{title}{I Give Up on {CHI}/{UIST}}.
\newblock \bibinfo{howpublished}{\url{http://dubfuture.blogspot.com/2009/11/i-give-up-on-chiuist.html}}.
\newblock
\newblock
\shownote{Blog post on the author’s personal blog ``DUB For the Future''}.


\bibitem[Liang et~al\mbox{.}(2022)]%
        {liang2022holistic}
\bibfield{author}{\bibinfo{person}{Percy Liang}, \bibinfo{person}{Rishi Bommasani}, \bibinfo{person}{Tony Lee}, \bibinfo{person}{Dimitris Tsipras}, \bibinfo{person}{Dilara Soylu}, \bibinfo{person}{Michihiro Yasunaga}, \bibinfo{person}{Yian Zhang}, \bibinfo{person}{Deepak Narayanan}, \bibinfo{person}{Yuhuai Wu}, \bibinfo{person}{Ananya Kumar}, {et~al\mbox{.}}} \bibinfo{year}{2022}\natexlab{}.
\newblock \showarticletitle{Holistic Evaluation of Language Models}.
\newblock \bibinfo{journal}{\emph{arXiv preprint arXiv:2211.09110}} (\bibinfo{year}{2022}).
\newblock
\href{https://doi.org/10.48550/arXiv.2211.09110}{doi:\nolinkurl{10.48550/arXiv.2211.09110}}


\bibitem[Ma et~al\mbox{.}(2024)]%
        {ma2024schrodinger}
\bibfield{author}{\bibinfo{person}{Zilin Ma}, \bibinfo{person}{Yiyang Mei}, \bibinfo{person}{Krzysztof~Z. Gajos}, {and} \bibinfo{person}{Ian Arawjo}.} \bibinfo{year}{2024}\natexlab{}.
\newblock \showarticletitle{Schr{\"o}dinger’s Update: User Perceptions of Uncertainties in Proprietary Large Language Model Updates}. In \bibinfo{booktitle}{\emph{Extended Abstracts of the {CHI} Conference on Human Factors in Computing Systems}}. \bibinfo{pages}{1--9}.
\newblock
\href{https://doi.org/10.1145/3613905.3651100}{doi:\nolinkurl{10.1145/3613905.3651100}}


\bibitem[Masson et~al\mbox{.}(2025)]%
        {masson2025textoshop}
\bibfield{author}{\bibinfo{person}{Damien Masson}, \bibinfo{person}{Young-Ho Kim}, {and} \bibinfo{person}{Fanny Chevalier}.} \bibinfo{year}{2025}\natexlab{}.
\newblock \showarticletitle{Textoshop: Interactions Inspired by Drawing Software to Facilitate Text Editing}. In \bibinfo{booktitle}{\emph{Proceedings of the 2025 {CHI} Conference on Human Factors in Computing Systems}} \emph{(\bibinfo{series}{CHI '25})}. \bibinfo{publisher}{Association for Computing Machinery}, \bibinfo{address}{New York, NY, USA}, Article \bibinfo{articleno}{1087}, \bibinfo{numpages}{14}~pages.
\newblock
\href{https://doi.org/10.1145/3706598.3713862}{doi:\nolinkurl{10.1145/3706598.3713862}}


\bibitem[Mitchell et~al\mbox{.}(2019)]%
        {mitchell2019model}
\bibfield{author}{\bibinfo{person}{Margaret Mitchell}, \bibinfo{person}{Simone Wu}, \bibinfo{person}{Andrew Zaldivar}, \bibinfo{person}{Parker Barnes}, \bibinfo{person}{Lucy Vasserman}, \bibinfo{person}{Ben Hutchinson}, \bibinfo{person}{Elena Spitzer}, \bibinfo{person}{Inioluwa~Deborah Raji}, {and} \bibinfo{person}{Timnit Gebru}.} \bibinfo{year}{2019}\natexlab{}.
\newblock \showarticletitle{Model Cards for Model Reporting}. In \bibinfo{booktitle}{\emph{Proceedings of the Conference on Fairness, Accountability, and Transparency}} \emph{(\bibinfo{series}{FAT* '19})}. \bibinfo{publisher}{Association for Computing Machinery}, \bibinfo{address}{New York, NY, USA}, \bibinfo{pages}{220--229}.
\newblock
\href{https://doi.org/10.1145/3287560.3287596}{doi:\nolinkurl{10.1145/3287560.3287596}}


\bibitem[Movva et~al\mbox{.}(2024)]%
        {movva2024topics}
\bibfield{author}{\bibinfo{person}{Rajiv Movva}, \bibinfo{person}{Sidhika Balachandar}, \bibinfo{person}{Kenny Peng}, \bibinfo{person}{Gabriel Agostini}, \bibinfo{person}{Nikhil Garg}, {and} \bibinfo{person}{Emma Pierson}.} \bibinfo{year}{2024}\natexlab{}.
\newblock \showarticletitle{Topics, Authors, and Institutions in Large Language Model Research: Trends from 17k {arXiv} Papers}. In \bibinfo{booktitle}{\emph{Proceedings of the 2024 Conference of the North American Chapter of the Association for Computational Linguistics: Human Language Technologies (Volume 1: Long Papers)}}. \bibinfo{pages}{1223--1243}.
\newblock
\href{https://doi.org/10.18653/v1/2024.naacl-long.67}{doi:\nolinkurl{10.18653/v1/2024.naacl-long.67}}


\bibitem[Olsen~Jr.(2007)]%
        {olsen2007usability}
\bibfield{author}{\bibinfo{person}{Dan~R. Olsen~Jr.}} \bibinfo{year}{2007}\natexlab{}.
\newblock \showarticletitle{Evaluating User Interface Systems Research}. In \bibinfo{booktitle}{\emph{Proceedings of the 20th Annual {ACM} Symposium on User Interface Software and Technology}} \emph{(\bibinfo{series}{UIST '07})}. \bibinfo{publisher}{Association for Computing Machinery}, \bibinfo{address}{New York, NY, USA}, \bibinfo{pages}{251--258}.
\newblock
\href{https://doi.org/10.1145/1294211.1294256}{doi:\nolinkurl{10.1145/1294211.1294256}}


\bibitem[Pang et~al\mbox{.}(2025)]%
        {pangLLMificationCHI2025}
\bibfield{author}{\bibinfo{person}{Rock~Yuren Pang}, \bibinfo{person}{Hope Schroeder}, \bibinfo{person}{Kynnedy~Simone Smith}, \bibinfo{person}{Solon Barocas}, \bibinfo{person}{Ziang Xiao}, \bibinfo{person}{Emily Tseng}, {and} \bibinfo{person}{Danielle Bragg}.} \bibinfo{year}{2025}\natexlab{}.
\newblock \showarticletitle{Understanding the {LLM}-ification of {CHI}: Unpacking the Impact of {LLMs} at {CHI} Through a Systematic Literature Review}. In \bibinfo{booktitle}{\emph{Proceedings of the 2025 {CHI} Conference on Human Factors in Computing Systems}} \emph{(\bibinfo{series}{CHI '25})}. \bibinfo{publisher}{Association for Computing Machinery}, \bibinfo{address}{New York, NY, USA}, Article \bibinfo{articleno}{456}, \bibinfo{numpages}{20}~pages.
\newblock
\href{https://doi.org/10.1145/3706598.3713726}{doi:\nolinkurl{10.1145/3706598.3713726}}


\bibitem[Park et~al\mbox{.}(2023)]%
        {park2023generative}
\bibfield{author}{\bibinfo{person}{Joon~Sung Park}, \bibinfo{person}{Joseph O’Brien}, \bibinfo{person}{Carrie~Jun Cai}, \bibinfo{person}{Meredith~Ringel Morris}, \bibinfo{person}{Percy Liang}, {and} \bibinfo{person}{Michael~S. Bernstein}.} \bibinfo{year}{2023}\natexlab{}.
\newblock \showarticletitle{Generative Agents: Interactive Simulacra of Human Behavior}. In \bibinfo{booktitle}{\emph{Proceedings of the 36th Annual {ACM} Symposium on User Interface Software and Technology}} \emph{(\bibinfo{series}{UIST '23})}. \bibinfo{publisher}{Association for Computing Machinery}, \bibinfo{address}{New York, NY, USA}, Article \bibinfo{articleno}{2}, \bibinfo{numpages}{22}~pages.
\newblock
\href{https://doi.org/10.1145/3586183.3606763}{doi:\nolinkurl{10.1145/3586183.3606763}}


\bibitem[Passi and Jackson(2018)]%
        {passi_trust_data_science}
\bibfield{author}{\bibinfo{person}{Samir Passi} {and} \bibinfo{person}{Steven~J. Jackson}.} \bibinfo{year}{2018}\natexlab{}.
\newblock \showarticletitle{Trust in Data Science: Collaboration, Translation, and Accountability in Corporate Data Science Projects}.
\newblock \bibinfo{journal}{\emph{Proceedings of the {ACM} on Human-Computer Interaction}} \bibinfo{volume}{2}, \bibinfo{number}{CSCW}, Article \bibinfo{articleno}{136} (\bibinfo{year}{2018}), \bibinfo{numpages}{28}~pages.
\newblock
\href{https://doi.org/10.1145/3274405}{doi:\nolinkurl{10.1145/3274405}}


\bibitem[Shaikh et~al\mbox{.}(2024)]%
        {shaikh2024rehearsal}
\bibfield{author}{\bibinfo{person}{Omar Shaikh}, \bibinfo{person}{Valentino~Emil Chai}, \bibinfo{person}{Michele Gelfand}, \bibinfo{person}{Diyi Yang}, {and} \bibinfo{person}{Michael~S. Bernstein}.} \bibinfo{year}{2024}\natexlab{}.
\newblock \showarticletitle{Rehearsal: Simulating Conflict to Teach Conflict Resolution}. In \bibinfo{booktitle}{\emph{Proceedings of the 2024 {CHI} Conference on Human Factors in Computing Systems}} \emph{(\bibinfo{series}{CHI '24})}. \bibinfo{publisher}{Association for Computing Machinery}, \bibinfo{address}{New York, NY, USA}, Article \bibinfo{articleno}{920}, \bibinfo{numpages}{20}~pages.
\newblock
\href{https://doi.org/10.1145/3613904.3642159}{doi:\nolinkurl{10.1145/3613904.3642159}}


\bibitem[Shaikh et~al\mbox{.}(2025)]%
        {shaikh2025gum}
\bibfield{author}{\bibinfo{person}{Omar Shaikh}, \bibinfo{person}{Shardul Sapkota}, \bibinfo{person}{Shan Rizvi}, \bibinfo{person}{Eric Horvitz}, \bibinfo{person}{Joon~Sung Park}, \bibinfo{person}{Diyi Yang}, {and} \bibinfo{person}{Michael~S. Bernstein}.} \bibinfo{year}{2025}\natexlab{}.
\newblock \showarticletitle{Creating General User Models from Computer Use}. In \bibinfo{booktitle}{\emph{Proceedings of the 38th Annual {ACM} Symposium on User Interface Software and Technology}} \emph{(\bibinfo{series}{UIST '25})}. \bibinfo{publisher}{Association for Computing Machinery}, \bibinfo{address}{New York, NY, USA}, Article \bibinfo{articleno}{35}, \bibinfo{numpages}{23}~pages.
\newblock
\href{https://doi.org/10.1145/3746059.3747722}{doi:\nolinkurl{10.1145/3746059.3747722}}


\bibitem[Shankar et~al\mbox{.}(2024)]%
        {shankar2024validates}
\bibfield{author}{\bibinfo{person}{Shreya Shankar}, \bibinfo{person}{J.~D. Zamfirescu-Pereira}, \bibinfo{person}{Bjoern Hartmann}, \bibinfo{person}{Aditya Parameswaran}, {and} \bibinfo{person}{Ian Arawjo}.} \bibinfo{year}{2024}\natexlab{}.
\newblock \showarticletitle{Who Validates the Validators? Aligning {LLM}-Assisted Evaluation of {LLM} Outputs with Human Preferences}. In \bibinfo{booktitle}{\emph{Proceedings of the 37th Annual {ACM} Symposium on User Interface Software and Technology}} \emph{(\bibinfo{series}{UIST '24})}. \bibinfo{publisher}{Association for Computing Machinery}, \bibinfo{address}{New York, NY, USA}, Article \bibinfo{articleno}{131}, \bibinfo{numpages}{14}~pages.
\newblock
\href{https://doi.org/10.1145/3654777.3676450}{doi:\nolinkurl{10.1145/3654777.3676450}}


\bibitem[Soden et~al\mbox{.}(2024)]%
        {soden2024evaluating}
\bibfield{author}{\bibinfo{person}{Robert Soden}, \bibinfo{person}{Austin Toombs}, {and} \bibinfo{person}{Michaelanne Thomas}.} \bibinfo{year}{2024}\natexlab{}.
\newblock \showarticletitle{Evaluating Interpretive Research in {HCI}}.
\newblock \bibinfo{journal}{\emph{Interactions}} \bibinfo{volume}{31}, \bibinfo{number}{1} (\bibinfo{year}{2024}), \bibinfo{pages}{38--42}.
\newblock
\href{https://doi.org/10.1145/3633200}{doi:\nolinkurl{10.1145/3633200}}


\bibitem[Sturdee et~al\mbox{.}(2021)]%
        {sturdee2021consequences}
\bibfield{author}{\bibinfo{person}{Miriam Sturdee}, \bibinfo{person}{Joseph Lindley}, \bibinfo{person}{Conor Linehan}, \bibinfo{person}{Chris Elsden}, \bibinfo{person}{Neha Kumar}, \bibinfo{person}{Tawanna Dillahunt}, \bibinfo{person}{Regan Mandryk}, {and} \bibinfo{person}{John Vines}.} \bibinfo{year}{2021}\natexlab{}.
\newblock \showarticletitle{Consequences, Schmonsequences! Considering the Future as Part of Publication and Peer Review in Computing Research}. In \bibinfo{booktitle}{\emph{Extended Abstracts of the 2021 {CHI} Conference on Human Factors in Computing Systems}} \emph{(\bibinfo{series}{CHI EA '21})}. \bibinfo{publisher}{Association for Computing Machinery}, \bibinfo{address}{New York, NY, USA}, Article \bibinfo{articleno}{95}, \bibinfo{numpages}{4}~pages.
\newblock
\href{https://doi.org/10.1145/3411763.3441330}{doi:\nolinkurl{10.1145/3411763.3441330}}


\bibitem[Subramonyam et~al\mbox{.}(2023)]%
        {subramonyam2023bridging}
\bibfield{author}{\bibinfo{person}{Hariharan Subramonyam}, \bibinfo{person}{Roy Pea}, \bibinfo{person}{Christopher~Lawrence Pondoc}, \bibinfo{person}{Maneesh Agrawala}, {and} \bibinfo{person}{Colleen Seifert}.} \bibinfo{year}{2023}\natexlab{}.
\newblock \showarticletitle{Bridging the Gulf of Envisioning: Cognitive Design Challenges in {LLM} Interfaces}.
\newblock \bibinfo{journal}{\emph{arXiv preprint arXiv:2309.14459}} (\bibinfo{year}{2023}).
\newblock
\href{https://doi.org/10.48550/arXiv.2309.14459}{doi:\nolinkurl{10.48550/arXiv.2309.14459}}


\bibitem[Syriani et~al\mbox{.}(2024)]%
        {syriani2024screeningarticles}
\bibfield{author}{\bibinfo{person}{Eugene Syriani}, \bibinfo{person}{Istvan David}, {and} \bibinfo{person}{Gauransh Kumar}.} \bibinfo{year}{2024}\natexlab{}.
\newblock \showarticletitle{Screening articles for systematic reviews with ChatGPT}.
\newblock \bibinfo{journal}{\emph{Journal of Computer Languages}}  \bibinfo{volume}{80} (\bibinfo{year}{2024}), \bibinfo{pages}{101287}.
\newblock
\showISSN{2590-1184}
\href{https://doi.org/10.1016/j.cola.2024.101287}{doi:\nolinkurl{10.1016/j.cola.2024.101287}}


\bibitem[Tanaka(2025)]%
        {tanaka2025assistedDeskReject}
\bibfield{author}{\bibinfo{person}{Yudai Tanaka}.} \bibinfo{year}{2025}\natexlab{}.
\newblock \bibinfo{title}{Revised {CHI} 2026 Papers Desk Reject Process}.
\newblock \bibinfo{howpublished}{\url{https://chi2026.acm.org/2025/08/08/revised-chi-2026-papers-desk-reject-process/}}.
\newblock
\newblock
\shownote{Blog post on the {CHI} 2026 Technical Program blog}.


\bibitem[Toyama(2015)]%
        {toyama2015geek}
\bibfield{author}{\bibinfo{person}{Kentaro Toyama}.} \bibinfo{year}{2015}\natexlab{}.
\newblock \bibinfo{booktitle}{\emph{Geek Heresy: Rescuing Social Change from the Cult of Technology}}.
\newblock \bibinfo{publisher}{PublicAffairs}.
\newblock


\bibitem[Vaithilingam et~al\mbox{.}(2025)]%
        {semantic_commit}
\bibfield{author}{\bibinfo{person}{Priyan Vaithilingam}, \bibinfo{person}{Munyeong Kim}, \bibinfo{person}{Frida-Cecilia Acosta-Parenteau}, \bibinfo{person}{Daniel Lee}, \bibinfo{person}{Amine Mhedhbi}, \bibinfo{person}{Elena~L. Glassman}, {and} \bibinfo{person}{Ian Arawjo}.} \bibinfo{year}{2025}\natexlab{}.
\newblock \showarticletitle{Semantic Commit: Helping Users Update Intent Specifications for AI Memory at Scale}. In \bibinfo{booktitle}{\emph{Proceedings of the 38th Annual ACM Symposium on User Interface Software and Technology}} \emph{(\bibinfo{series}{UIST '25})}. \bibinfo{publisher}{Association for Computing Machinery}, \bibinfo{address}{New York, NY, USA}, Article \bibinfo{articleno}{137}, \bibinfo{numpages}{18}~pages.
\newblock
\showISBNx{9798400720376}
\href{https://doi.org/10.1145/3746059.3747778}{doi:\nolinkurl{10.1145/3746059.3747778}}


\bibitem[Weyssow et~al\mbox{.}(2022)]%
        {syriani2022recommending}
\bibfield{author}{\bibinfo{person}{Martin Weyssow}, \bibinfo{person}{Houari Sahraoui}, {and} \bibinfo{person}{Eugene Syriani}.} \bibinfo{year}{2022}\natexlab{}.
\newblock \showarticletitle{Recommending metamodel concepts during modeling activities with pre-trained language models}.
\newblock \bibinfo{journal}{\emph{Software and Systems Modeling}} \bibinfo{volume}{21}, \bibinfo{number}{3} (\bibinfo{year}{2022}), \bibinfo{pages}{1071--1089}.
\newblock


\bibitem[Wobbrock and Kientz(2016)]%
        {wobbrock2016research}
\bibfield{author}{\bibinfo{person}{Jacob~O. Wobbrock} {and} \bibinfo{person}{Julie~A. Kientz}.} \bibinfo{year}{2016}\natexlab{}.
\newblock \showarticletitle{Research Contributions in Human--Computer Interaction}.
\newblock \bibinfo{journal}{\emph{Interactions}} \bibinfo{volume}{23}, \bibinfo{number}{3} (\bibinfo{year}{2016}), \bibinfo{pages}{38--44}.
\newblock
\href{https://doi.org/10.1145/2907069}{doi:\nolinkurl{10.1145/2907069}}


\bibitem[Yang et~al\mbox{.}(2020)]%
        {yang2020re}
\bibfield{author}{\bibinfo{person}{Qian Yang}, \bibinfo{person}{Aaron Steinfeld}, \bibinfo{person}{Carolyn Ros{\'e}}, {and} \bibinfo{person}{John Zimmerman}.} \bibinfo{year}{2020}\natexlab{}.
\newblock \showarticletitle{Re-examining Whether, Why, and How Human-{AI} Interaction Is Uniquely Difficult to Design}. In \bibinfo{booktitle}{\emph{Proceedings of the 2020 {CHI} Conference on Human Factors in Computing Systems}}. \bibinfo{pages}{1--13}.
\newblock
\href{https://doi.org/10.1145/3313831.3376301}{doi:\nolinkurl{10.1145/3313831.3376301}}


\bibitem[Zhang and Arawjo(2025)]%
        {zhang2025chainbuddy}
\bibfield{author}{\bibinfo{person}{Jingyue Zhang} {and} \bibinfo{person}{Ian Arawjo}.} \bibinfo{year}{2025}\natexlab{}.
\newblock \showarticletitle{{ChainBuddy}: An {AI}-Assisted Agent System for Generating {LLM} Pipelines}. In \bibinfo{booktitle}{\emph{Proceedings of the 2025 {CHI} Conference on Human Factors in Computing Systems}} \emph{(\bibinfo{series}{CHI '25})}. \bibinfo{publisher}{Association for Computing Machinery}, \bibinfo{address}{New York, NY, USA}, Article \bibinfo{articleno}{241}, \bibinfo{numpages}{21}~pages.
\newblock
\href{https://doi.org/10.1145/3706598.3714085}{doi:\nolinkurl{10.1145/3706598.3714085}}


\bibitem[Zheng et~al\mbox{.}(2025)]%
        {zheng2025evalignux}
\bibfield{author}{\bibinfo{person}{Qingxiao Zheng}, \bibinfo{person}{Minrui Chen}, \bibinfo{person}{Pranav Sharma}, \bibinfo{person}{Yiliu Tang}, \bibinfo{person}{Mehul Oswal}, \bibinfo{person}{Yiren Liu}, {and} \bibinfo{person}{Yun Huang}.} \bibinfo{year}{2025}\natexlab{}.
\newblock \showarticletitle{{EvAlignUX}: Advancing {UX} Evaluation Through {LLM}-Supported Metrics Exploration}. In \bibinfo{booktitle}{\emph{Proceedings of the 2025 {CHI} Conference on Human Factors in Computing Systems}} \emph{(\bibinfo{series}{CHI '25})}. \bibinfo{publisher}{Association for Computing Machinery}, \bibinfo{address}{New York, NY, USA}, Article \bibinfo{articleno}{1051}, \bibinfo{numpages}{25}~pages.
\newblock
\href{https://doi.org/10.1145/3706598.3714045}{doi:\nolinkurl{10.1145/3706598.3714045}}


\bibitem[Zhou et~al\mbox{.}(2025)]%
        {zhou2025instructpipe}
\bibfield{author}{\bibinfo{person}{Zhongyi Zhou}, \bibinfo{person}{Jing Jin}, \bibinfo{person}{Vrushank Phadnis}, \bibinfo{person}{Xiuxiu Yuan}, \bibinfo{person}{Jun Jiang}, \bibinfo{person}{Xun Qian}, \bibinfo{person}{Kristen Wright}, \bibinfo{person}{Mark Sherwood}, \bibinfo{person}{Jason Mayes}, \bibinfo{person}{Jingtao Zhou}, \bibinfo{person}{Yiyi Huang}, \bibinfo{person}{Zheng Xu}, \bibinfo{person}{Yinda Zhang}, \bibinfo{person}{Johnny Lee}, \bibinfo{person}{Alex Olwal}, \bibinfo{person}{David Kim}, \bibinfo{person}{Ram Iyengar}, \bibinfo{person}{Na Li}, {and} \bibinfo{person}{Ruofei Du}.} \bibinfo{year}{2025}\natexlab{}.
\newblock \showarticletitle{InstructPipe: Generating Visual Blocks Pipelines with Human Instructions and {LLMs}}. In \bibinfo{booktitle}{\emph{Proceedings of the 2025 {CHI} Conference on Human Factors in Computing Systems}} \emph{(\bibinfo{series}{CHI '25})}. \bibinfo{publisher}{Association for Computing Machinery}, \bibinfo{address}{New York, NY, USA}, Article \bibinfo{articleno}{877}, \bibinfo{numpages}{22}~pages.
\newblock
\href{https://doi.org/10.1145/3706598.3713905}{doi:\nolinkurl{10.1145/3706598.3713905}}


\bibitem[Zimmerman et~al\mbox{.}(2007)]%
        {zimmerman2007research}
\bibfield{author}{\bibinfo{person}{John Zimmerman}, \bibinfo{person}{Jodi Forlizzi}, {and} \bibinfo{person}{Shelley Evenson}.} \bibinfo{year}{2007}\natexlab{}.
\newblock \showarticletitle{Research Through Design as a Method for Interaction Design Research in {HCI}}. In \bibinfo{booktitle}{\emph{Proceedings of the SIGCHI Conference on Human Factors in Computing Systems}}. \bibinfo{pages}{493--502}.
\newblock
\href{https://doi.org/10.1145/1240624.1240704}{doi:\nolinkurl{10.1145/1240624.1240704}}


\end{thebibliography}

\appendix 

\section{Connections Between Findings and Considerations}

In Table~\ref{tab:traceability_matrix}, we make explicit how each reporting consideration relates to subsections of our findings, for readers interested in how we derived our initial considerations and suggestions from the findings. Note that section 4.3.15 reviews some considerations that were explicitly suggested by multiple participants; some of our considerations reflect these, especially in 5.3.

\section{Examples of How Reporting is Mediated by Centrality of LLM Components to Claims} \label{appendix-examples}

To clarify how the centrality of LLM components to author claims mediates the expected rigor of reporting efforts, we provide two examples, drawing from two real published papers at CHI and UIST. 

First, consider Rehearsal, an LLM-integrated system that authors claim helps novice conflict negotiators learn conflict negotiation skills~\cite{shaikh2024rehearsal}. The LLM components---a chatbot to simulate human behavior and targeted feedback to help novices improve---are central to author's claims that \textit{the system helps novices realistically rehearse and learn conflict negotiation skills}. The problem domain of conflict negotiation is also  sensitive and high-stakes. The paper should, and does, include a modest technical evaluation outside of a user study to validate and quantify the robustness of the LLM components in relation to these claims. Authors also report system failure modes, spend considerable space describing their prompting pipeline, and sketch out a prompt template which operationalizes a key theoretical framework they employ called Interests-Rights-Power.

By contrast, consider Textoshop, a paper exploring the interface metaphor of Photoshop for text editing, with ``words as pixels, sentences as regions, and tones as colours''~\cite{masson2025textoshop}. Authors' primary contribution is \textit{introducing and exploring a novel interaction design,} with prompting taking a secondary, enabling role to testing the concept, without claiming robustness. Implementation details for specific tools, like joining two texts together, appear swappable to non-LLM techniques without affecting this main claim---i.e., author's design contribution seems to hold regardless of near-future advancements in AI models, the exact details of prompts, or even the fine-grained details of the performance of individual components. Because it would have little impact on readers' ability to validate this claimed contribution, authors do not report a technical evaluation. Authors justify LLM usage in the context of how their system helps users overcome challenges of natural language prompting, and briefly report a more involved prompting methodology for one tool, but leave the remaining transparency over LLM aspects to open-sourced code.  %
Finally, authors also lightly de-center LLMs/AI in paper framing, not mentioning the term LLM or AI in their title and abstract.

\begin{table*}[h!]
\centering
\caption{Mapping Reporting Guidelines to Empirical Findings}
\label{tab:traceability_matrix}
\Description{Table with two columns, showing guidelines in the left column, and the findings section number(s) it relates to in the second column.}
\renewcommand{\arraystretch}{1.3} %
\begin{tabular}{|p{0.3\linewidth}|p{0.6\linewidth}|}
\toprule
\textbf{Proposed Guideline} & \textbf{Grounding in Empirical Findings (Section 4)} \\
\midrule

\textbf{Transparency regarding prompts and configuration} & 
\textbf{Finding 4.2.1:} Participants argued against ``reporting all prompts,'' favoring selective reporting based on how critical the prompt is to the claim, how difficult it was to engineer, or whether others could learn from it. \newline
\textbf{Finding 4.2.2:} Authors use appendices to manage complexity and length. \\

\textbf{Visualizing system architecture and data flow} & 
\textbf{Finding 4.3.8:} Reviewers expressed a desire for the ``gist'' of the workflow rather than exhaustive text; participants recommended diagrams/tables to show data flow. \newline
\textbf{Finding 4.2.3:} Authors use ``sketches'' of architecture to manage complexity. \\

\textbf{Technical validation of LLM components central to claims} & 
\textbf{Finding 4.3.4:} 14 participants perceived that technical evaluations are becoming expected to counter skepticism. \newline
\textbf{Findings 4.1.3-4:} Uncertainty of LLMs distinguishes them from code-based systems, requiring trust-building to build confidence in system performance. \\

\textbf{Transparency Regarding Failure Modes of LLM components} & 
\textbf{Finding 4.3.15:} Multiple participants (P4, P6, P8, P11-12, P14, P16) explicitly suggested guidelines for reporting mistakes/potential harms. \newline
\textbf{Finding 4.1.4:} Authors noted the importance of identifying failure modes during iteration. \\

\textbf{Justifying the choice to use LLMs} & 
\textbf{Finding 4.3.9:} Ten participants noted reviewers expect justification for LLM use. \newline
\textbf{Finding 4.1.2:} Authors choose proprietary models for specific practical reasons (latency, cost, capabilities) rather than pure performance. \\

\textbf{Documenting engineering processes of LLM components} & 
\textbf{Finding 4.3.3:} Conveying care and thought that went into engineering LLM components can improve trust with reviewers. \newline
\textbf{Finding 4.1.4:} The iterative engineering of LLM components is a key and necessary part of their development, yet authors can fail to report these processes. \\

\textbf{Framing the contribution to HCI beyond the application of LLMs} & 
\textbf{Finding 4.2.4:} Authors have developed a strategy of de-emphasizing ``LLM/AI'' in titles, abstracts, or introductions to avoid reviewer bias and focus on the HCI contribution. \newline
\textbf{Finding 4.3.1:} Fear of ``knee-jerk reactions'' from reviewers regarding AI. \\

\textbf{Future-proofing contributions against future AI advancement} & 
\textbf{Finding 4.3.2:} Authors seek to future-proof contributions with strategies like ``our novelty is interaction design'' to remain relevant regardless of model capability. \newline
\textbf{Finding 4.3.13:} Reviewers dismiss non-LLM components with ``can't an AI do that?'' \\

\bottomrule
\end{tabular}
\end{table*}

\end{document}